\documentclass[useAMS,usenatbib]{mnras}

\usepackage{amsmath}
\usepackage{amssymb}
\usepackage{txfonts}
\usepackage{graphicx}

\newcommand{\Msun}{\ensuremath{\mathrm{M}_\odot}}
\newcommand{\ud}{\ensuremath{\mathrm{d}}}
\newcommand{\pd}{\ensuremath{\partial}}
\newcommand{\hf}{\ensuremath{\frac{1}{2}}}
\newcommand{\dt}{\ensuremath{\Delta t}}
\newcommand{\ka}{\ensuremath{\kappa_\mathrm{a}}}
\newcommand{\ks}{\ensuremath{\kappa_\mathrm{s}}}

\graphicspath{{./}{figures/}}

\title[A Novel Boltzmann Transport Scheme]{A Novel Multi-Dimensional Boltzmann Neutrino Transport Scheme for Core-Collapse Supernovae}

\author[Chan \& M\"uller]{
Conrad~Chan\thanks{conrad.chan@monash.edu}
and 
Bernhard~M\"uller\thanks{bernhard.mueller@monash.edu}
\\
School of Physics and Astronomy,
Monash University, Clayton, Australia, VIC 3800
}

\begin{document}

\label{firstpage}
\pagerange{\pageref{firstpage}--\pageref{lastpage}}
\maketitle

\begin{abstract}
We introduce a new discrete-ordinate scheme for solving the general relativistic Boltzmann transport equation in the context of core-collapse supernovae.  Our algorithm avoids the need to spell out the complicated advection terms in energy and angle that arise when the transport equation is formulated in spherical polar coordinates, in the comoving frame, or in a general relativistic spacetime. We instead approach the problem by calculating the advection of neutrinos across momentum space using an intuitive particle-like approach that has excellent conservation properties and fully accounts for Lorentz boosts, general relativistic effects, and grid geometry terms. In order to avoid the need for a global implicit solution, time integration is performed using a locally implicit Lax-Wendroff scheme that correctly reproduces the diffusion limit. This will facilitate the use of our method on massively parallel distributed-memory architectures. We have verified the accuracy and stability of our scheme with a suite of test problems in spherical symmetry and axisymmetry. To demonstrate that the new algorithm works stably in core-collapse supernova simulations, we have coupled it to the general relativistic hydrodynamics code \textsc{CoCoNuT} and present a first demonstration run of a $20\Msun$ progenitor with a reduced set of neutrino opacities.
\end{abstract}


\section{Introduction} \label{sec:intro}

Since \cite {1966ApJ...143..626C} proposed that neutrinos play a crucial role in reviving the stalled shock in core-collapse supernovae (CCSNe), numerical models have incorporated neutrino transport in some form, with varying degrees of sophistication. The neutrino transport problem in CCSNe poses one of the most computationally challenging problems to date. The extreme range in densities within the supernova results in optically thick regions in which neutrinos are diffusive, and optically thin regions in which neutrinos are decoupled from the fluid and can stream freely. The so-called ``gain region'' behind the stalled shock, which the success of the explosion hinges upon, lies within the semi-transparent region between the two regimes, where an accurate treatment of the neutrino distribution function is important \citep[for modern reviews of the topic, see][]{2005ARNPS..55..467M,2012ARNPS..62..407J,2013RvMP...85..245B}. The most accurate treatment of neutrino transport calls for the solution of the full Boltzmann equation \citep{1984oup..book.....M}, 

\begin{equation}
    \frac{\partial f}{\partial t} + \mathbf{u} \cdot \nabla f = \mathcal{C}(f),
    \label{eq:boltzmann}
\end{equation}

\begin{figure}
    \centering
    \includegraphics[width=\linewidth]{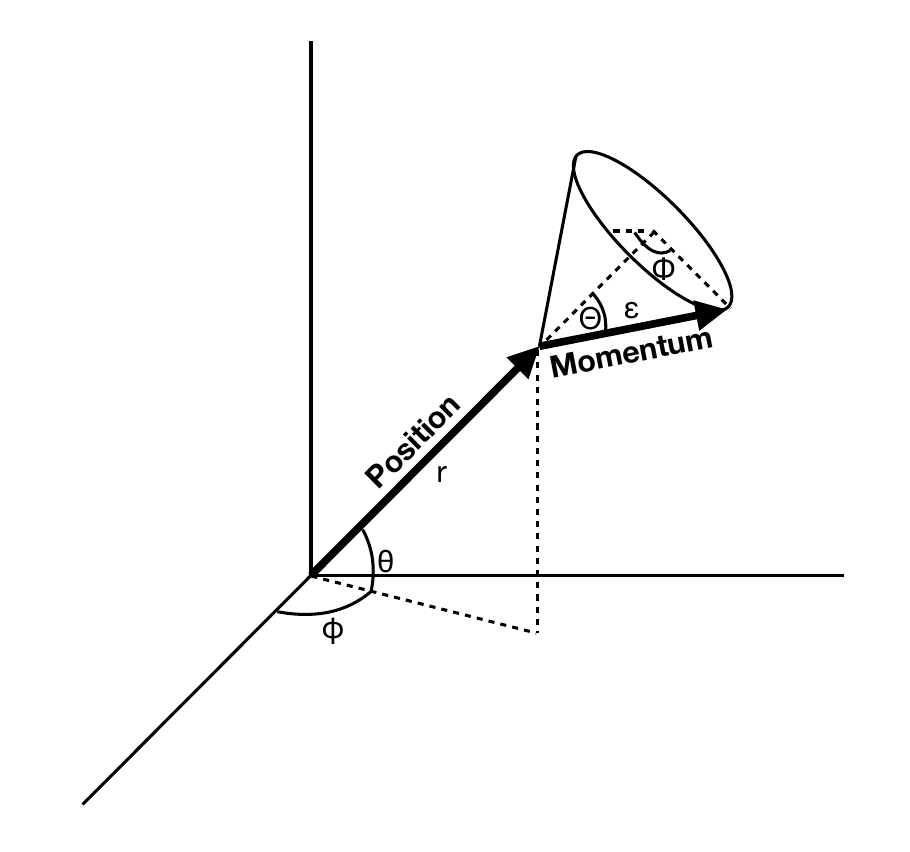}
    \caption{Illustration of the six-dimensional phase space. The coordinates $r$, $\theta$, and $\phi$ specify the position of a neutrino, while $\Theta$ and $\Phi$ specify the direction of propgation, with $\epsilon$ specifying the energy (magnitude). The $\Theta$ angle is conventionally expressed as $\mu=\cos\Theta$, where $\mu\in\left[-1,1\right]$.
    }
    \label{fig:diagram_angles}
\end{figure}

(shown here the Newtonian form for clarity) which describes the neutrino momentum distribution $f$ at every point in space (Figure \ref{fig:diagram_angles}). Evolving through time, this results in a 7-dimensional problem. The left-hand side of the equation is the advection equation, which couples together the spatial dimensions. The right-hand side contains the collision integral, which accounts for interactions between neutrinos and matter, and couple together the momentum space dimensions. Interaction rates are sensitive to the neutrino energy, necessitating an energy-dependent treatment. Additionally, there are three flavors of neutrinos (electron, muon, and tauon) and respective antineutrinos for a total of six species. Due to the different interaction rates of each species, their transport behaviour is considerably different, thus the core-collapse problem also calls for a flavour-dependent treatment. In general, even flavour can evolve via neutrino oscillations, a purely quantum mechanical effect \citep{1978PhRvD..17.2369W,1985YaFiz..42.1441M}.

The first serious attempts to incorporate neutrino transport into core-collapse simulations relied on the flux-limited diffusion (FLD) scheme in spherical symmetry, one of the simplest approximations of radiation transport \citep{1970ApJ...160..959A,1977ApJ...218..815A,1982ApJS...50..115B,1985ApJS...58..771B,1987ApJ...318..744M,1989ApJ...339..978B,1992ApJ...398..531C}. 
 FLD
is still employed in several modern
supernova codes as an economical
approach to multi-dimensional transport
\citep[e.g.][]{2005ApJ...626..317W,2018arXiv180905608B,2019MNRAS.490.3545R}.
This approximation solves the diffusion equation for mean intensity of radiation with the assumption that the angular distribution is isotropic, discarding directional dependence of the field (while in ``multi-group'' implementations still retaining energy dependence). In the limit of high optical depth, the diffusion approximation is the exact solution. At low optical depth where neutrinos stream freely, a flux-limiter is necessary to prevent advection above the maximum transport velocity (i.e. the speed of light). The semi-transparent region where shock revival is decided by neutrino heating, however, is handled using ad-hoc switching prescriptions. Since the choice of flux-limiter determines the flux-factor, which in turn affects the neutrino heating rate, this  can lead to inaccurate results \citep{1999A&A...344..533Y,2000ApJ...539..865B}. Comparisons with more sophisticated schemes have shown that the FLD approximation smears out anisotropies in the radiation field \citep{2008ApJ...685.1069O}. This is especially crucial in rotating models, which exhibit a large-scale asymmetry in the post-shock flow. Nonetheless, these approximations were necessary in the pioneering CCSN simulations due to the sheer computational infeasability of solving the full Boltzmann equation at the time.

That is not to say that schemes to solve the full Boltzmann equation had not yet been devised. The discrete ordinate method ($S_N$) had already been developed by \cite{Carlson1967} to solve the \emph{neutron} transport problem in nuclear reactors, which was later adapted to supernova modelling \citep{1977ApJ...217..565Y,1993ApJ...410..740M}. The $S_N$ method is a brute-force approach, directly discretising all of the variables of the Boltzmann equation, in particular the direction of radiation propagation. In practice, this is extremely computationally expensive, so the first CCSN simulations to apply this scheme were run in spherical symmetry, which reduced the dimensionality of the problem \citep{1999A&A...344..533Y,2001PhRvL..86.1935M,2001PhRvD..63j3004L,2004ApJS..150..263L}.

Methods based on solving the moment equations derived from the Boltzmann equation (so called ``moment methods'') have also been  employed. In these methods, the individual directions of radiation are integrated over to obtain the moments of radiation, which become the variables of the Boltzmann equation. An arbitrary number of equations can be derived, though the solution of each equation requires the solution for a higher moment, thus requiring the system to be truncated at some point using an approximate closure relation obtained from the known moments. Truncating at the zeroth moment yields the diffusion approximation, while modern schemes are most commonly two-moment methods, solving for energy and momentum. The closure may be an algebraic relation \citep[e.g.][]{1978JQSRT..20..541M,1984JQSRT..31..149L}, as implemented by \citet{2000MNRAS.317..550P,2015MNRAS.453.3386J,2015ApJS..219...24O,2016ApJS..222...20K,2016ApJ...831...98R,2016ApJ...831...81S,2019ApJS..241....7S}. The use of an algebraic closure provides significant computational cost savings while still being sufficiently accurate in spherical symmetry \citep{2017ApJ...847..133R}.
Alternatively, the closure can be obtained by solving a model Boltzmann equation using the moments in variable Eddington factor (VEF) methods \citep{2000ApJ...539L..33R,2000ApJ...539..865B,2002A&A...396..361R}. Moment methods model the angular dependence of radiation while avoiding direct discretisation of the momentum space. Implementations that use a closure based on an exact solution of the Boltzmann equation are still considered as Boltzmann transport solvers.

As the field evolved to the understanding that multi-dimensional fluid flows are important in supernovae, there too became a demand for multi-dimensional neutrino transport solvers. Computational power, however, could not keep up, and compromises were made in most multi-dimensional simulations. Early attempts at multi-dimensional calculations discarded energy-dependence entirely \citep{1994ApJ...435..339H,1995ApJ...450..830B,2000ApJ...541.1033F,2004ApJ...601..391F}, though these ``gray'' simulations have since been shown to yield considerably different results compared their energy-dependent counterparts.  Implicit finite differencing, which was employed in many spherically symmetric solutions, is expensive in two and three dimensions, so multi-group codes often resort to the ``ray-by-ray'' approximation and solves the transport problem in one dimension along angular rays and assuming the neutrino distribution to be axially symmetric around the radial direction \citep{2006A&A...457..281B}. Others have continued to use the multi-group flux-limited diffusion approximation \citep{2005ApJ...626..317W,2007ApJ...664..416B,2009ApJS..181....1S,2019MNRAS.490.3545R}, which can also
be combined with the
ray-by-ray approximation \citep{2018arXiv180905608B}. Meanwhile, truly multidimensional $S_N$ schemes were developed \citep{2004ApJ...609..277L,2012ApJS..199...17S,2014ApJS..214...16N} , although some omitting energy redistribution and velocity-dependent effects to avoid the implicit coupling across energy groups and similarly, true multidimensional moment schemes were also developed  \citep{2007ApJ...659.1458H,2015MNRAS.453.3386J,2016ApJS..222...20K,2016ApJ...831...98R,2016ApJ...831...81S,2019ApJS..241....7S}. 

There are a handful of other ambitious treatments that have been explored, for example, the $P_N$ scheme of \cite{2013JCoPh.242..648R} which decomposes the momentum space into spherical harmonics, and the spectral scheme of \cite{2014CQGra..31d5012P} which decomposes real space using a Fourier basis and momentum space using a Chevyshev basis. These approaches, however,
still remain in the demonstration phase
and have not yet been used for full supernova
simulations.
As an alternative approach to solving the Boltzmann equation, the Monte Carlo (MC) method, in which the trajectories of a large sample of particles are directly evolved, solves the transport problem with full angular dependence \citep{1971JCoPh...8..313F,1984JCoPh..54..508F}. It has been applied in the context of CCSN simulations \citep{1989A&AS...78..375J,1992A&A...256..452J,2012ApJ...755..111A,2017ApJ...847..133R} and it has also been used to supplement the two-moment scheme by providing a closure \citep{2018MNRAS.475.4186F}. Though MC methods are  accurate and scale excellently, they are generally noisy, requiring high resolution for a smooth solution. Recently, the particle-like moment closure method of \citet{2020ApJ...891..118R} has demonstrated improved convergence compared to MC, though it is yet to be applied to neutrino transport.

To enable more affordable multi-dimensional simulations in the interim as we await for advancements in computing power, there
has also been an interest in alternative efficient transport
approximations in recent years.
\cite{2009ApJ...698.1174L} have made improvements to the diffusion approximation, and \cite{2015MNRAS.448.2141M} have devised a fast approximation to the stationary transport equation
using a combination of a two-stream solution of the Boltzmann equation and an algebraic Eddington factor to obtain closure.

Aside from limits in computing power, there are several challenges on the algorithm front. Most importantly, a numerical solution to the Boltzmann equation should be stable, while accurately reproducing the correct behaviour in the diffusion and free-streaming limit. These requirements have led to the proliferation of moment methods, which satisfy physical limits by construction, and are easier to discretise in a numerically stable way. Current schemes that solve the Boltzmann equation rely on a global implicit solution to ensure stability. 
A traditional explicit implementation has timesteps severely restricted by the Courant limit of neutrinos travelling at the speed of light, commonly an order of magnitude smaller than that of the fluid. Most problematically, the source terms on the right-hand-side of the equation are stiff and non-linear, necessitating an even smaller time-step for stability, which an implicit scheme circumvents.

The major downside of fully implicit schemes is that they necessitate an iterative solution that couples the entire grid, usually in real space \emph{and} momentum space. Achieving good parallel scaling with this degree of coupling is difficult. While in spherical symmetry it is possible to fit the entire domain on a single processor, 2D and 3D simulations often span multiple computer nodes, usually decomposed based on the spatial grid. Since an implicit solution requires information from the entire grid at every timestep, inter-node communication becomes a bottleneck. While the ray-by-ray scheme circumvents this by implicitly solving the 1D solution along rays, this is not true multi-dimensional transport. 
While the ray-by-ray method
may yield acceptable results in supernova
simulations of non-rotating progenitors
\citep{2019ApJ...873...45G},
errors may be significant in the presence of large-scale asymmetries in the radiation field,
for example when dealing with rapidly rotating neutron stars, accretion disks, and jet outflows.

In this paper, we describe a three-dimensional $S_N$ scheme to solve the Boltzmann equation in general relativity, implemented explicitly in real space to allow scalability but implicitly in momentum space to ensure stability. This explicit-implicit hybrid approach \citep{ASCHER1997151} has been used in the context of neutrino transport by \citet{2013JCoPh.242..648R,2015MNRAS.453.3386J,2019JCoPh.389...62C}. The space-explicit implementation has the benefit of easier maintenance and the flexibility to be extended to cover additional physics, e.g. treating neutrino oscillations by solving the quantum kinetic equations \citep{1993NuPhB.406..423S,2008PhRvD..78h5017C,2015IJMPE..2441009V,2019PhRvD..99l3014R}. Although in an explicit scheme, timesteps are restricted by the Courant condition, the sound speed in the proto-NS interior is a significant fraction of the neutrino velocity ($c$), so the neutrino timesteps and fluid timesteps are not too dissimilar. We postulate that for large multi-dimensional simulations, the improved scaling of our scheme will make up for the additional cost of shorter timesteps. An explicit scheme also opens up the possibility for variable time-stepping based on the local Courant limit, an optimisation that is practically impossible in an implicit scheme. We couple our scheme to the \textsc{CoCoNuT} hydrodynamics solver \citep{2005PhRvD..71f4023D,2010ApJS..189..104M} with the conformal flatness condition (CFC) for the spacetime \citep{2008IJMPD..17..265I,1996PhRvD..54.1317W,2009PhRvD..79b4017C}. Just as the first solutions to the Boltzmann equation were developed well before they were affordable, the exascale computing power required to use our scheme in 3D is still in the distant horizon. We demonstrate the capabilities of our scheme in 1D and 2D, and argue that it will be equally applicable in future 3D simulations.

\section{Solution Strategy}

\subsection{A ``derivative-free'' approach}
\label{sec:approach}
The left-hand side of the Boltzmann equation is simply an advection equation. It is preferred, however, to solve it in the comoving frame, since the collision terms on the right-hand side is naturally expressed in the comoving frame.
The Lorenz transformation of the collision integral from the comoving frame to the
lab frame is cumbersome, and achieving the correct diffusion limit in a lab-frame
approach is non-trivial. These problems are somewhat mitigated in the
mixed-frame approach where the lab frame opacities are expressed in terms of
the comoving frame opacities by means of a first-order Taylor expansion,
but this approach is inherently limited to $\mathcal{O}(v/c)$ in accuracy.
There is, however, a downside to the comoving frame approach, since it results in advection across the momentum dimensions of the phase space, in the form of energy and angle derivatives that can obscure the original meaning of the advection equation. In this section, we explain how these terms come about and propose a strategy for avoiding them.

For problems with spherical geometry, radiation typically transitions towards a radially-peaked distribution. On such a grid, a natural and commonly adopted momentum space coordinate is one that is aligned with the radial direction in real space. The complication, however, is that the direction of propagation with respect to the radial direction of a particle with a non-radial momentum component varies as it travels inwards or outwards. This manifests as a $\frac{\pd}{\pd \mu}$ derivative when the Boltzmann equation is expressed in the spherically symmetric form
\begin{equation}
\frac{1}{c} \frac{\pd f}{\pd t} 
+ \mu \frac{\pd f}{\pd r}
+ \frac{1-\mu^2}{r} \frac{\pd f}{\pd \mu}
= \mathfrak{C}\left[f\right].
\end{equation}

When a moving background fluid is introduced, the Boltzmann equation becomes even more complicated in the comoving frame. As a particle moves from one cell to another, the neutrino energy and direction of propagation shifts, manifesting as the $\frac{\pd}{\pd \epsilon}$ term and an additional $\frac{\pd}{\pd \mu}$ term respectively, and an additional $\frac{\pd f}{\pd r}$ term due to the advection of neutrinos with the fluid, giving the $\mathcal{O}(v/c)$ accurate equation
\begin{align}
\nonumber
    \frac{1}{c}\frac{\pd f}{\pd t} 
    &+ \left(\mu + \beta\right) \frac{\pd f}{\pd r} 
    + \frac{1-\mu^2}{r}\frac{\pd f}{\pd \mu}
    + \left(1-\mu^2\right)\left[\mu\left(\frac{\beta}{r}-\frac{\pd \beta}{\pd r}\right)
    - \frac{1}{c}\frac{\pd \beta}{\pd t}\right]\frac{\pd f}{\pd \mu}
    \\
    &
    +
    \mu\epsilon \left[ \mu\left(\frac{\beta}{r} - \frac{\pd \beta}{\pd r}\right) - \frac{1}{c}\frac{\pd \beta}{\pd t} - \frac{1}{\mu}\frac{\beta}{r}\right] \frac{\pd f}{\pd \epsilon}
    = \mathfrak{C}\left[f\right]
\end{align}
where $\beta = v / c$ \citep{1984oup..book.....M}. Once a curved spacetime is included, the Boltzmann equation quickly becomes unwieldy, with the introduction of many additional derivatives that account for gravitational ray-bending and redshift:
\begin{equation}
    p^{\hat{\mu}} L^{\mu}_{\ \hat{\mu}} \frac{\pd f}{\pd x^\mu} - \Gamma^{\hat{\imath}}_{\ \hat{\nu}\hat{\mu}} p^{\hat{\nu}} p^{\hat{\mu}} P^{\tilde{\imath}}_{\ \hat{\imath}} \frac{\pd f}{\pd p^{\tilde{\imath}}} = \mathfrak{C}\left[f\right]
\end{equation}
where the $x^\mu$ are the spacetime position coordinates in the lab frame, $p^{\tilde{\imath}}$ are the momentum space coordinates in the comoving frame, $\Gamma^{\hat{\imath}}_{\ \hat{\nu}\hat{\mu}}$ are the Christoffel symbols, $L^{\mu}_{\ \hat{\mu}}$ is the coordinate transformation from the lab frame to the comoving frame, and $P^{\tilde{\imath}}_{\ \hat{\imath}}$ is the transformation between curvilinear and cartesian momentum space coordinates \cite[see][for a derivation]{2013PhRvD..88b3011C}. To implement the equation into a numerical scheme, each of the terms have to be converted into a finite difference representation, with care required to maintain any desired conservation properties \citep[e.g. the $S_N$ scheme of][]{2004ApJS..150..263L}.

All of these derivatives, however, arise simply as a consequence of the change of basis in momentum space as a particle is transported from one cell to another and as the metric and velocity field evolve in time. One can eliminate some of the derivatives by expressing the equation in the lab frame, but the derivatives associated with the grid geometry will remain, and the collision term becomes much more complicated because the opacities need to be transformed. 

Rather than implementing each derivative explicitly in the Boltzmann equation, we instead propose in our solution to transport neutrinos along characteristics (geodesics), handling the advection terms in the lab frame while keeping the distribution function in the lab frame. In their destination cell, advected neutrinos are mapped back to the
momentum space grid in the comoving frame based on their comoving-frame four-velocity.
This strategy is reminiscent of the semi-Lagrangian method for advection problems, only
that the mapping of the advected solution is carried out in momentum space only.

\begin{figure*}
    \centering
    \includegraphics[width=0.7\textwidth]{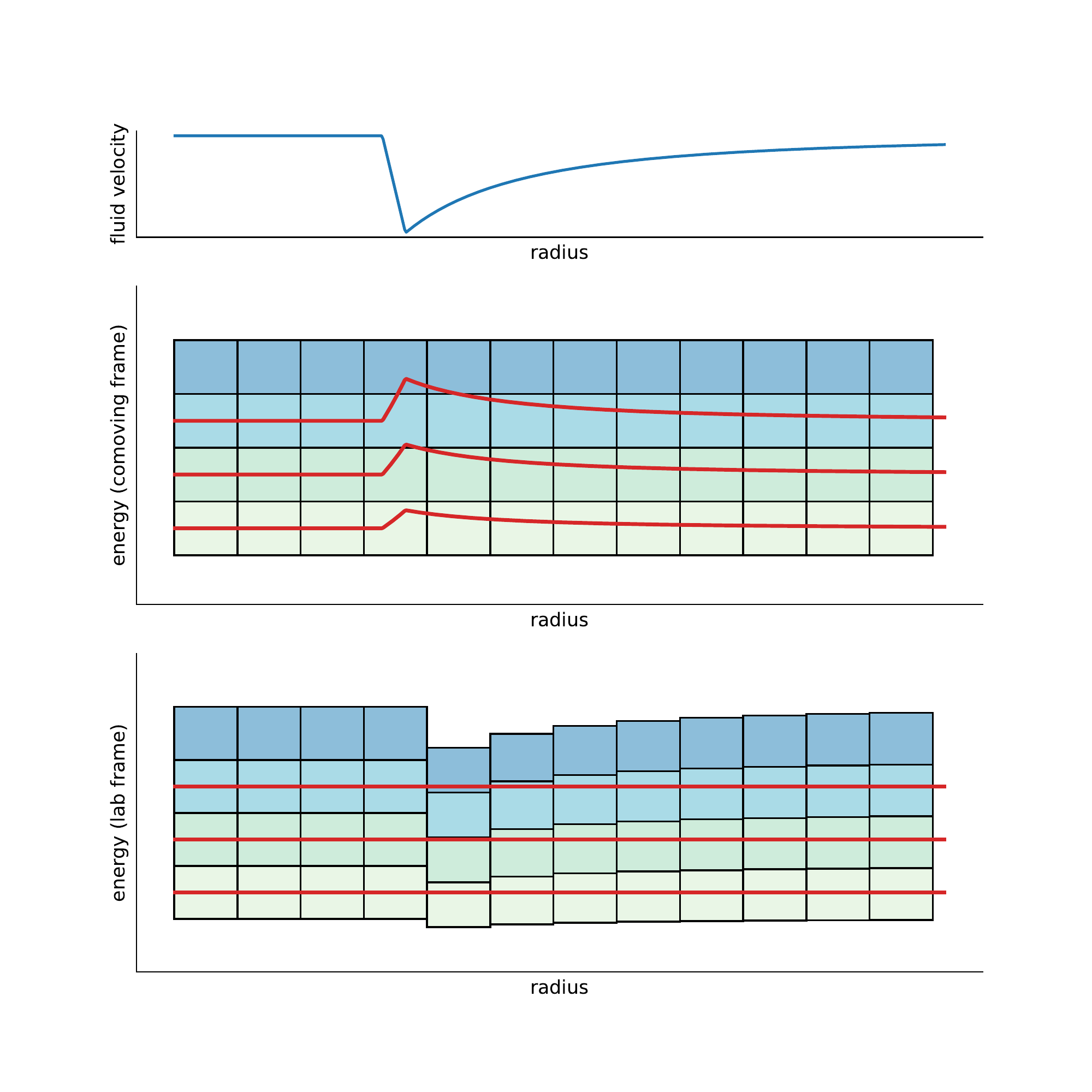}
    \caption{Illustration of the advection in energy space as a neutrino moves through the typical shock velocity profile (top), as viewed in the comoving frame (center) and the lab frame (bottom). Red curves indicate the characteristics of neutrinos through this slice of the phase space as they travel outwards. The shaded cells represent the finite-volume elements of the scheme.
    }
    \label{fig:diagram_energy}
\end{figure*}

We illustrate our idea in Figure \ref{fig:diagram_energy} by considering neutrinos streaming outwards through a fluid moving inwards with a typical shock velocity profile. Observers comoving with the shock will observe the neutrino energy to be blue-shifted. In traditional discretisations of the Boltzmann equation, this advection of neutrinos to a higher energy group is handled by the $\frac{\pd}{\pd \epsilon}$ derivative. We instead consider representing the advection of neutrinos in the lab frame, where the neutrino energy is constant but the energy groups in each cell are shifted relative to their neighbours. The key to our method is that the blue-shift in neutrino energy is accounted for by fluxes through these misaligned interfaces. No derivatives need to appear in our discretisation --- they are instead handled automatically by the misalignment of cells. This procedure can also be applied to the transition of propagation angle to $\mu\rightarrow1$ as neutrinos move outwards. In fact, it can be applied to any transformation between bases in momentum space. The remaining task is then to identify the appropriate transformation between the bases of each cell, or in terms of our illustration, to find the fluxes through the misaligned interfaces, so that a neutrino originating from one cell can be mapped into the basis of the destination cell in a conservative manner, which we explain in detail in Section \ref{subsec:destinationmapping}.

Another strategy to avoid derivatives in energy is to use different energy space grids in the lab frame and in the comoving frame \citet{2014ApJS..214...16N}. In terms of implementation, our single (comoving) frame approach is in fact conceptually similar in that we also employ linear interpolation to map between adjacent frames. Where our scheme differs is that we map directly to the destination frame, bypassing the intermediate lab frame.\footnote{The scheme of \citet{2014ApJS..214...16N} also differs in that it is \emph{globally} implicit, coupling together space and momentum dimensions.} Our implementation goes beyond this existing work by applying the remapping procedure generally, to not only the shifts in energy but also angle, arising from not only Lorentz boosts but also geometric rotations and general relativistic Hamiltonian ``kicks''.

\subsection{Treatment of relativistic effects} \label{subsec:greffects}
General relativistic (GR) effects become important in core-collapse supernovae after the formation of the neutron star, and indeed, most state-of-the-art models account for these effects either with GR transport \citep{1999A&A...344..533Y,2004ApJS..150..263L,2010ApJS..189..104M,2016ApJ...831...98R,2016ApJS..222...20K}, or with an approximate GR treatment using Newtonian pseudopotentials \citep{2002A&A...396..361R,2006A&A...445..273M}. Our scheme has been developed with the intention of coupling to GR hydrodynamic codes. For simplicity, we restrict ourselves to the CFC case, but our scheme is readily generalised to GR simulations without this approximation.

\begin{figure*}
    \centering
    \includegraphics[width=\textwidth]{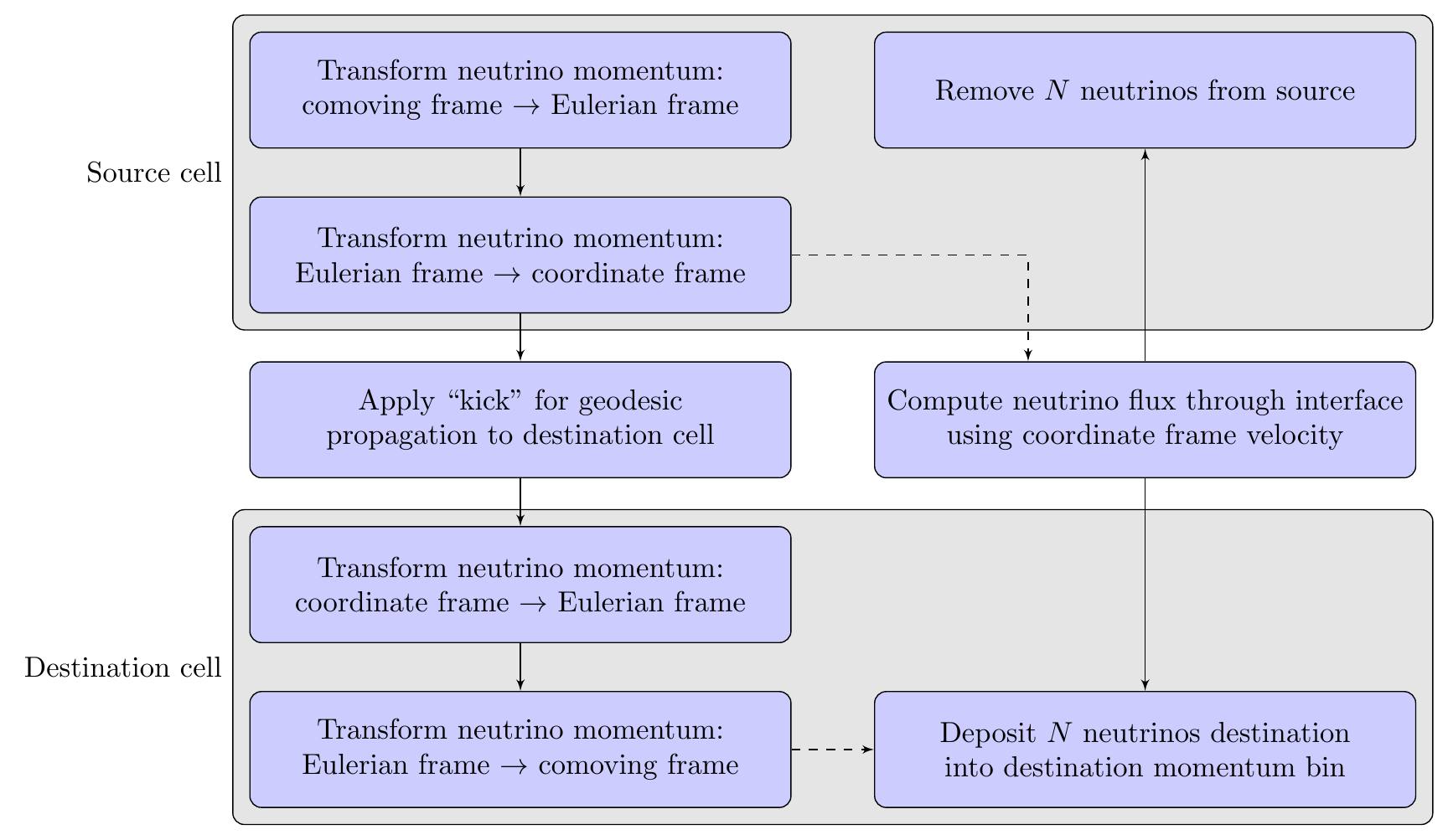}
    \caption{Flowchart of the chain of transformations from the source comoving frame to the destination comoving frame. The boxes on the right show how the upwind flux (Section \ref{subsec:upwind}) is computed. The fluxes are calculated in the coordinate frame, and the neutrinos are deposited into the destination comoving frame in the destination momentum bin obtained from the chain of transformations. The procedure for the Lax-Wendroff flux (Section \ref{subsec:lw}) is conceptually the same if we treat the reconstructed interface distribution as the destination cell for the half-steps, and use the reconstructed coordinate-frame velocity for the flux computation in the full step.}
    \label{fig:flowchart}
\end{figure*}

When a neutrino moves in space and time, its lab-frame velocity components will be deflected according to the geodesic equation. The lab-frame velocity is calculated by first transforming the comoving-frame velocity into the Eulerian frame using the Lorentz boost, and then into the lab frame using the metric. After the deflection is calculated, the comoving-frame velocity in the destination cell is calculated by applying the frame transformations in reverse. The fluxes through cell interfaces (Section \ref{subsec:fluxes}) are always calculated using the lab-frame velocities. The full chain of transformations  from the source cell (src) to the destination cell (dst) is
\begin{equation}
    \tilde{u}_\mathrm{src} \xrightarrow{\Lambda_\mathrm{src}} \hat{u}_\mathrm{src} \xrightarrow{M_\mathrm{src}} u_\mathrm{src} \xrightarrow{G,K} u_\mathrm{dst} \xrightarrow{M_\mathrm{dst}^{-1}} \hat{u}_\mathrm{dst} \xrightarrow{\Lambda_\mathrm{dst}^{-1}} \tilde{u}_\mathrm{dst}.
    \label{eq:transformchain}
\end{equation}
where $\tilde{u}$, $\hat{u}$, and $u$ are the four-velocities in the comoving, Eulerian, and coordinate frame. The Lorentz transformation $\tilde{u}\leftrightarrow\hat{u}$ is described in Section \ref{subsubsec:lorentz}. The transformation $\hat{u}\leftrightarrow u$ to and from the coordinate frame is described in Section \ref{subsubsec:frametransform}. The transformation $u_\mathrm{src}\rightarrow u_\mathrm{dst}$ accounting for the geometric rotation and Ricci rotation of velocity components between cells is described in Sections \ref{subsubsec:geometric} and \ref{subsubsec:ricci}. We show how each of these velocities are used in our scheme in Figure (\ref{fig:flowchart}).

\subsection{Reproducing the diffusion limit using a stable explicit scheme}
A finite-difference scheme is convergent if and only if it is consistent with the original equation and it is stable. It is essential for any Boltzmann scheme to be able to reproduce the correct solution within all regimes encountered during core-collapse. While reproducing the advection equation in the optically thin limit is relatively trivial, the optically thick limit requires more care. In optically thick regions where the mean free path of a neutrino of order $\mathcal{O}(\mathrm{cm})$ is much smaller than the zone width, the scheme should yield the correct diffusion limit, which even the most primitive flux-limited diffusion schemes achieve by construction. In principle, this challenge can be avoided by choosing a sufficiently fine grid spacing that resolves the mean free path of neutrinos, but in practice, it is exceedingly computationally expensive to do so. The more prudent solution is to use a differencing scheme that represents a valid discretisation of the diffusion equation.

The formulation of \cite{Carlson1967} solves the problem by using a backward-time centered-space scheme, which is consistent with the diffusion limit. When \cite{1977ApJ...217..565Y} applied this scheme to the core-collapse problem, however, they found that oscillatory behaviour arises when the distribution is close to kinetic equilibrium or in the presence of steep gradients. To circumvent this, they instead adopt upwind differencing in space. Although stable, the step method generally produces incorrect fluxes in the diffusion limit \citep{1989JCoPh..83..212L}, so they introduce an artificial diffusion term. On the other hand, \cite{1993ApJ...405..669M} avoid the need for a diffusion term in their implementation by interpolating between centered differencing and upwind differencing (i.e. between first- and second-order differencing in space) to achieve consistency in optically thick regions and avoid oscillations in problematic regions.

To achieve stability, all of these schemes have used implicit backward-time differencing, which is unconditionally stable. Though this circumvents a restrictive $\Delta t \lesssim \frac{\kappa \Delta x^2}{c}$ limit (where $\kappa$ is the opacity) on the timestep, an accurate solution in the regime where advection is dominant (i.e. the P\'eclet number $\mathrm{Pe}\sim\kappa\Delta x$ is small) still requires timesteps that are not too much greater than 
the stability limit
$\Delta t=\Delta x/c$ for explicit advection. Explicit implementations of high-resolution shock-capturing schemes are not suitable, usually because they are unstable. Any reconstruction steps that maintain stability will introduce jumps at interfaces, leading to numerical diffusion that disrupts convergence to the physical diffusion limit. In moment schemes, convergence to the diffusion limit is usually achieved by an ad-hoc treatment \citep{2002astro.ph..6281A,2007A&A...464..429G,2015ApJS..219...24O}. These approaches are not satisfactory for our purposes, since they require the diffusion limit to be known. In GR and when solving the more general quantum-kinetic equations, the diffusion limit is non-trivial.\footnote{
Very few method papers for recent
diffusion or two-moment codes for core-collapse
supernovae actually address the issue of the
correct diffusion limit in detail. \citep[e.g.][]{2019MNRAS.490.3545R}.  However,
in addition to including the correct metric
factors when computing spatial and temporal
derivatives,
the correct diffusive flux in 
a (nearly) static GR space-time
must be computed using spatial derivatives
of the zeroth moment $J$ at constant
neutrino energy at infinity, not
at 
constant comoving frame energy. 
The  diffusive flux must therefore generally include
an additional energy derivative
of the zeroth-moment, unless an adapted
energy grid in the comoving frame
is used as in \citep{2015MNRAS.448.2141M,2018arXiv180905608B}.
This is necessary to ensure that
the diffusive flux establishes
the correct thermal equilibrium
$\alpha T=\mathrm{const.}$
in a static space-time
\citep[e.g.][]{1930PhRv...36.1791T,1999ApJ...513..780P}.
} It is desirable to avoid manual intervention if possible.

We propose discretising using a locally-implicit (semi-explicit) Lax-Wendroff scheme, where the fluxes are computed using a half-timestep at the cell interfaces:
\begin{equation}
    \frac{f_{a+\hf}^{n+\hf} - f_{a+\hf}^n}{\frac{1}{2} \Delta t} + u\cdot\frac{f_{a+1}^n - f_a}{\Delta x} = \mathfrak{C}\left[f_{a+\hf}\right], \label{eq:lw}
\end{equation}
where $f_{a+\hf}^n = \frac{1}{2}\left(f_a^n + f_{a+1}^n\right)$. We show that this is consistent with the diffusion limit in Appendix \ref{app:diffusion}. The explicit time differencing of the advective fluxes has the advantage of avoiding a global coupling of the spatial grid, allowing computational scalability. The key to our approach is to use implicit time differencing for the collision terms at the half- and full-steps of the Lax-Wendroff scheme which allows the timestep $\Delta t$ to be independent of opacity $\kappa$ while coupling the grid only in momentum space. The Lax-Wendroff scheme is stable for the advection equation when the CFL limit is met, and the addition of a collision term will only further dampen the growth of errors. Similar to \cite{1993ApJ...405..669M}, we avoid oscillatory behaviour in optically thin cells by interpolating between Lax-Wendroff fluxes and upwind fluxes.

\section{Implementation}

\subsection{Discretisation} \label{subsec:discretisation}

We exploit the spherical symmetries of CCSNe by adopting spherical polar spatial coordinates
\begin{align}
    x &= r \sin\theta\cos\phi, \\
    y &= r \sin\theta\sin\phi, \\
    z &= r \cos\theta.
\end{align}
We discretise the six-dimensional phase space into finite volume elements. In real space, cell interfaces are defined as the surface on a plane bounded by the four line segments connecting four adjacent vertices of the grid. The vertices have coordinates
\begin{alignat}{2}
    r_{a+\hf} & \in \left[0, r_\mathrm{max}\right],\ && a = 0,\dots,N_r \\
    \theta_{b+\hf} & \in \left[\theta_\mathrm{min}, \theta_\mathrm{max}\right],\ && b = 0,\dots,N_\theta \\
    \phi_{c+\hf} & \in \left[\phi_\mathrm{min}, \phi_\mathrm{max}\right],\ && c = 0,\dots,N_\phi
\end{alignat}
where $r_{a+\hf}$ are spaced non-uniformly (arbitrary choice), and $\theta_{b+\hf}$ and $\phi_{c+\hf}$ are spaced uniformly.\footnote{We use the ``$i+\hf$'' notation to refer to the interface between cells $i$ and $i+1$.} The area $\Delta A$ of the interface is defined as the area of the flat surface bounded by the quadrilateral formed by four adjacent vertices, and the volume $\Delta V$ of the cell is defined as the volume bounded by the hexahedron (frustum) formed by eight adjacent vertices.\footnote{These hexahedra are not the same as the wedges of a spherical polar grid, as all their faces are flat.}

We also adopt spherical coordinates for the momentum space:
\begin{align}
    p_x &= \epsilon \sin\Theta\cos\Phi, \\
    p_y &= \epsilon \sin\Theta\sin\Phi, \\
    p_z &= \epsilon \cos\Theta.
\end{align}
We use the definition $\mu=\cos\Theta$. At each point on the grid, the $\mu=1$ momentum cell is aligned with the radial direction (Figure \ref{fig:diagram_angles}), to exploit the fact that the distribution transitions to radially-peaked at large radii. This means the momentum space vectors at two points with differing angular coordinates are not the same. We discretise the momentum vector into spherical shells bounded by the surfaces
\begin{equation}
    \epsilon_{i+\hf} \in \left[\epsilon_\mathrm{min}, \epsilon_\mathrm{max}\right], i = 0,\dots,N_\epsilon
\end{equation}
and solid angle elements $\Delta\Omega$ bounded by the coordinates
\begin{alignat}{2}
    \mu_{j+\hf} & \in \left[-1, 1\right],\ && j = 0,\dots, N_\mu \\
    \Phi_{k+\hf} & \in \left[0, 2\pi\right],\ && k = 0,\dots,N_\epsilon
\end{alignat}
where $\mu_{j+\hf}$ are spaced such that the cell volumes are the Gauss-Lobatto weights \citep{Boyd2000}, and $\Phi_{k+\hf}$ are spaced uniformly. 

In the comoving frame, neutrinos have four-velocity
\begin{equation}
    u^\nu = (1, u^i),
\end{equation}
where $u^i$ is the three-velocity.\footnote{Neutrinos become important to the dynamics of core-collapse supernovae at MeV energy scales, at which their velocities are close to the speed of light $c$. Thus we may employ the massless approximation of neutrinos, i.e. approximating their velocity as $c$ such that their velocities satisfy $\|u^\nu\|=0$.} Using our chosen momentum space coordinates, the velocity of a neutrino is given by the zenith coordinate $\mu$ and the azimuthal angle $\Phi$, such that
\begin{equation}
    u^\nu = (1, \mu, \sqrt{1-\mu^2}\cos\Phi, \sqrt{1-\mu^2}\sin\Phi).
\end{equation}

The four-momentum vector of a neutrino is given by
\begin{equation}
    p^\nu = \epsilon u^\nu,
\end{equation}
where $\epsilon$ is the energy of the neutrino.

These three dimensions in momentum space and three dimensions in real space constitute the six-dimensional phase space. The distribution of neutrinos in this phase space is given by $f(r,\theta,\phi,\epsilon,\mu,\Phi)$. The four-current of particles in the phase space cell $\ud \Gamma_\mathrm{com}$ is given by
\begin{equation}
\label{eq:current}
  j^\mu=
\int f  u^\mu \,\ud \Gamma_\mathrm{com}.
\end{equation}
The number $\Delta N$ of particles in $\ud \Gamma_\mathrm{com}$ within some 3-volume is 
\begin{align}
\Delta N &=
  \int  \sqrt{-g}\,
  \langle \mathbf{j},
  \ud x^\mu\rangle \wedge \ud x^\nu \wedge \ud x^\xi \wedge \ud x^\sigma
  \nonumber
  \\
  &=\int  \sqrt{-g}  j^\mu\, \ud x^\nu\, \ud x^\xi\, \ud x^\sigma,
\end{align}
where $\langle\mathbf{j},dx^\mu\rangle$ denotes the contraction of the 4-vector $\mathbf{j}$ with the volume form. This general form can be used to obtain both the number of particles within a space-like three-volume cell $\ud^3x$ as well as the number of particles flowing through cell interfaces per coordinate time $\ud t$ (two space dimensions and one time dimension). The number of particles within the phase space volume
$\ud \Gamma_\mathrm{com}$
in a cell $\ud x^1 \,\ud x^2 \,\ud x^3$ is
\begin{equation}
N = \iiiint \sqrt{-g} f  u^0 \,\ud \Gamma_\mathrm{com}\, \ud x^1 \,\ud x^2 \,\ud x^3,
\end{equation}
and the flux across a cell interface with $x_1=\mathrm{const.}$ is
\begin{equation}
F = \iiiint \sqrt{-g} f u^1 \,\ud \Gamma_\mathrm{com}\,\ud x^2 \,\ud x^3 \,\ud t.
\end{equation}
For a conformally flat metric \citep{2008IJMPD..17..265I} (used in \textsc{CoCoNuT}), with
\begin{equation}
  g_{\mu\nu}=
  \left(
  \begin{array}{cccc}
    -\alpha^2+\beta_i \beta^i & \beta_x & \beta_y & \beta_z \\
    \beta_x & \phi^4 & 0 & 0\\
    \beta_y & 0 & \phi^4 & 0\\
    \beta_z & 0 & 0 &\phi^4 \\    
    \end{array}
  \right),
\end{equation}
in terms of the lapse function $\alpha$, the conformal factor $\phi$, the shift vector $\beta^i$, and $\beta_i=\phi^4 \beta^i$, these can be simplified to
\begin{align}
N &= 
\iiiint \phi^6 f  \hat{u}^0 \,\ud \Gamma_\mathrm{com}\,\ud x^1 \,\ud x^2 \,\ud x^3,\\
F &= 
\iiiint \phi^6 f \left(\frac{\alpha}{\phi^2}\hat{u}^1+\beta^r \hat{u}^0\right) \,\ud \Gamma_\mathrm{com}\, \ud x^2 \,\ud x^3 \,\ud t.
\end{align}
in terms of the Eulerian velocity components $\hat{u}^\mu$.
It is critical to recognise that because
of the invariance of the distribution function we can directly use
$f$ in the \emph{comoving} frame to compute the number
of neutrinos within phase space cell $\ud \Gamma_\mathrm{comv}$
that cross a zone boundary during the time interval $\ud t$.
Only the neutrino four-velocity in Equation~(\ref{eq:current})
must be transformed into the coordinate frame.
The problem, however, is that the current $j^\mu$ only
obeys a conservation law $\nabla_\mu j^\mu=0$ if we define
$\ud \Gamma_\mathrm{com}$ away from a given reference point
$x^\mu$ as the phase space volume filled by particles
from  $\ud \Gamma_\mathrm{com}$ at  $x^\mu$ (or in technical
terms if we define $\ud \Gamma_\mathrm{com}$ away from
$x^\mu$ by an exponential map). In general the exponential
map of  $\ud \Gamma_\mathrm{com}(x^\mu)$ from
$x^\mu $ to another point $x'^\mu $ will \emph{differ} from
the momentum space volume $\ud \Gamma_\mathrm{com}(x'^\mu)$ 
defined by the same comoving-frame momentum space coordinates,
which necessitates the remapping step that we outlined
in Section~\ref{sec:approach} and will specify in detail
in the next section.

\subsection{Mapping to destination cells} \label{subsec:destinationmapping}

\begin{figure}
    \centering
    \includegraphics[width=\linewidth]{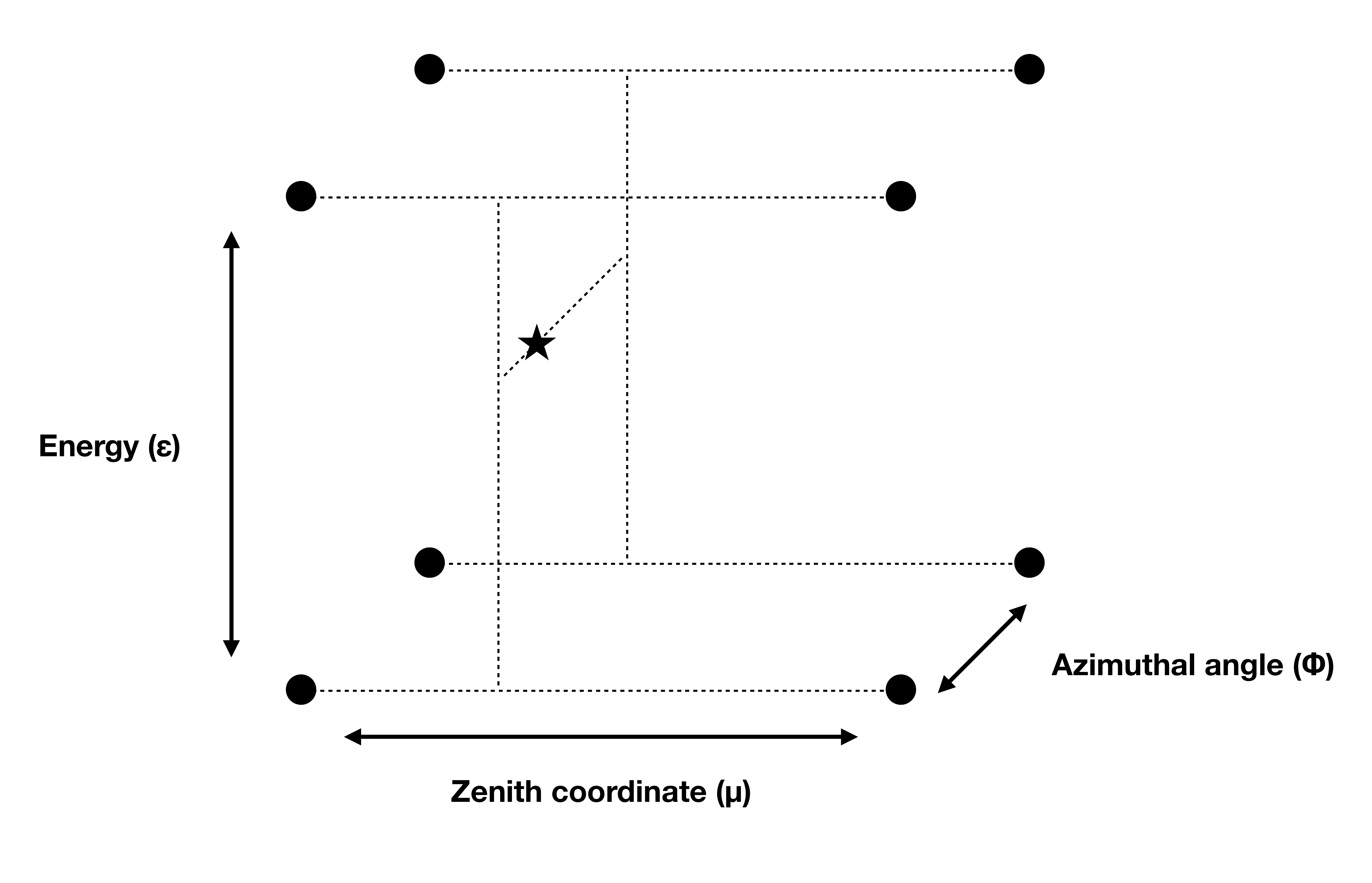}
    \caption{Illustration of the interpolation of an arriving momentum vector (star) as the linear combination of the nearest 8 vectors in the basis of the destination cell (dots). This method conserves energy and propagation angle of the net momentum vector by construction.}
    \label{fig:diagram_interpolation}
\end{figure}

As explained in Section \ref{sec:approach} the momentum coordinate of a spatially advected particle in its destination cell will not be the same as in the cell it originated from. Rather than handling these shifts using derivatives in the Boltzmann equation, we deposit the arriving neutrino into a linear combination of the nearest eight cells in momentum space (Figure~\ref{fig:diagram_interpolation}). As a neutrino arrives in a spatial cell with momentum $p_*=(\epsilon_*,\mu_*,\Phi_*)$ after it undergoes the full chain of transformations
\begin{equation}
    \tilde{u}_\mathrm{dst} = \Lambda^{-1}_\mathrm{dst}M^{-1}_\mathrm{dst}G(K(M_\mathrm{src}\Lambda_\mathrm{src}\tilde{u}_\mathrm{src}))
\end{equation}
(the formal expression of Equation \ref{eq:transformchain}), it is split between the two nearest cells in each coordinate in momentum space with weights such that both the energy, number, and direction of the neutrino are conserved, which amounts to a trilinear interpolation.

In order to write down finite-difference equations
for the entire scheme, it will be useful to formally interpret the
momentum-dependent distribution function as a vector 
$\mathbf{f}$ with components $f_{ijk}$. We can then
encode the remapping procedure as a linear operator $T$ that acts
on the source distribution function $\mathbf{f}_\mathrm{src}$
to yield $\mathbf{f}_\mathrm{dst}$,
\begin{equation}
\label{eq:transformation}
    \mathbf{f}_\mathrm{dst}=
    T[\mathbf{f}_\mathrm{src}],
\end{equation}
or in component form,
\begin{equation}
f^\mathrm{dst}_{i'j'k'}=
\sum_{ijk}
T_{i'j'k',ijk} f^\mathrm{src}_{ijk},
\end{equation}
Let us now consider a single \emph{column} of $T$ containing
the interpolation weights for neutrinos from phase space
cell $(i,j,k)$ into the eight destination cells in momentum
space that bracket the four-vector $\tilde{u}_\mathrm{dst}$
in the destination cell
(Figure~\ref{fig:diagram_interpolation}). The only non-zero
matrix elements in this column are those for the cells
immediately to the left (L)
or right (R) of $\tilde{u}_\mathrm{dst}$ with
$i'=i_\mathrm{L/R}$, $j'=j_\mathrm{L/R}$, and $k'=k_\mathrm{L/R}$.
 For example, $i_\mathrm{L}$ and $i_\mathrm{R}$
 are the two adjacent indices that fulfil
 $\epsilon_{i_\mathrm{L}} \leq \epsilon_* \leq \epsilon_{i_\mathrm{R}}$.
To obtain
the correct trilinear interpolation weights, we
factor these matrix elements into interpolation weights
$w_{i_\mathrm{L/R}}$, $w_{j_\mathrm{L/R}}$, and $w_{k_\mathrm{L/R}}$
for the $\epsilon$-, $\mu$-, and $\Phi$- direction,
\begin{equation}
T_{i_\mathrm{L/R} j_\mathrm{L/R}k_\mathrm{L/R},ijk}=w_{i_\mathrm{L/R}}w_{j_\mathrm{L/R}}w_{k_\mathrm{L/R}}.
\end{equation}
    \label{eq:inversetransform}
Requiring neutrino number conservation leads to the condition,
\begin{equation}
    \sum_{i' = i_\mathrm{L},i_\mathrm{R}} \sum_{j' = j_\mathrm{L},j_\mathrm{R}} \sum_{k' = k_\mathrm{L},k_\mathrm{R}} w_{i'}w_{j'}w_{k'} f_{ijk}^\mathrm{src}
    =
    f^\mathrm{src}_{ijk}.
    \label{eq:interpolation}
\end{equation}
To fulfil this condition, we choose the weights  such that
\begin{align}
        w_{i_\mathrm{L}} + w_{i_\mathrm{R}} &= 1, \\ 
        w_{j_\mathrm{L}} + w_{j_\mathrm{R}} &= 1, \\ w_{k_\mathrm{L}} + w_{k_\mathrm{R}} &= 1. 
\end{align}
Similarly, to conserve the direction of the neutrinos, the weights need to fulfill
\begin{align}
    \mu_* &= \mu_\mathrm{L} w_{j_\mathrm{L}} + \mu_\mathrm{R} w_{j_\mathrm{R}}, 
    \\
     \cos(\Phi_*) &= \cos(\Phi_\mathrm{L} w_{k_\mathrm{L}}) + \cos(\Phi_\mathrm{R} w_{k_\mathrm{R}}), 
     \\
     \sin(\Phi_*) &= \sin(\Phi_\mathrm{L} w_{k_\mathrm{L}}) + \sin(\Phi_\mathrm{R} w_{k_\mathrm{R}}),
\end{align}
where $\mu_\mathrm{L} \leq \mu_* \leq \mu_\mathrm{R}$ and $\Phi_\mathrm{L} \leq \Phi_* \leq \Phi_\mathrm{R}$. To conserve energy of the neutrinos, the weights need to fulfill
\begin{align}
    \epsilon_* = \epsilon_\mathrm{L} w_{i_\mathrm{L}} + \epsilon_\mathrm{R} w_{i_\mathrm{R}},
\end{align}
where $\epsilon_\mathrm{L} \leq \epsilon_* \leq \epsilon_\mathrm{R}$. 

This mapping procedure forms the core of our scheme, and used to connect cells in different points in space. It requires only the transformation of a momentum vector between two frames to be known, circumventing the need to manually implement the special and general relativistic derivative terms in the Boltzmann equation. 

This procedure is also applied as particles travel forward in time. As the metric and background fluid velocity evolves, we simply remap the distribution function in each cell at the beginning of each timestep using Equation (\ref{eq:transformation}), calculating $M_\mathrm{src}$ and $\Lambda_\mathrm{src}$ using values from the previous timestep, $M_\mathrm{dst}$ and $\Lambda_\mathrm{dst}$ using values from the current timestep, and $K$ using the metric from the previous and current timestep. Since particles are not travelling across the spatial grid, the geometric rotations do not apply, but relativistic boosts to their velocity can still cause them to move between angular bins.

While we implement our scheme on a spherical polar grid, this remapping procedure allows it to be adapted to any arbitrary grid geometry with minimal modifications. In a traditional implementation with derivatives, this would require major re-writing of code.

We now need to work out the precise form of the
Lorentz transform $\Lambda$, the transform $M$ from the Eulerian frame to the coordinate frame, and the transformation operators $G$ and $K$ for the geometric rotation and Hamiltonian ``kick'' between two points in space.

\subsubsection{Geometric rotations} \label{subsubsec:geometric}
When calculating fluxes between adjacent cells in the angular direction, the coordinates frames are rotated since the momentum space coordinates are aligned with the radial direction. The three-velocity of the arriving neutrino, after all relativistic boosts have been performed, must be rotated by an angle in the direction opposite to the direction of travel. Since this only needs to be performed for adjacent cells, the frames are only ever rotated in the $\theta$- or $\phi$-direction, but not both simultaneously.

The rotation of an arbitrary three-vector $\mathbf{u}$ about a unit vector $\mathbf{k}$ by angle $\beta$ is given by
\begin{equation}
    \mathbf{u}' = \cos\beta\mathbf{u} + \sin\beta\left(\mathbf{k}\times\mathbf{u}\right) + \left(\mathbf{k}\cdot\mathbf{u}\right)\left(1-\cos\beta\right)\mathbf{k}.
\end{equation}
The zeroth component of the four-velocity is unaffected
by the rotation, so the operator $G$ acting on the
four-vector  $(u^0,\mathbf{u})$ is just
\begin{equation}
    G(u^0,\mathbf{u}) = (u^0,\cos\beta\mathbf{u} + \sin\beta\left(\mathbf{k}\times\mathbf{u}\right) + \left(\mathbf{k}\cdot\mathbf{u}\right)\left(1-\cos\beta\right)\mathbf{k}).
\end{equation}

The shift necessary for a neutrino travelling in the $\theta$-direction is a rotation by $\beta=-\Delta\theta$ around $\mathbf{k} = (0,0,1)$ in the momentum space coordinates. Similarly, the shift necessary for a neutrino travelling in the $\phi$ direction is a rotation by $\beta=-\Delta\phi$ around $\mathbf{k}=(\sin\theta,\cos\theta,0)$.

\subsubsection{Lorentz boosts}
\label{subsubsec:lorentz}
In our scheme, the neutrino source terms are always computed in the comoving frame, and similarly, the momentum grid is also defined in the comoving frame. Where the fluid velocity is non-zero, the neutrino velocity must first be transformed into the Eulerian frame to the calculate fluxes across the cell interfaces, and then transformed into the comoving frame of the destination cell to determine the momentum bin that the neutrino arrives into. 

The boosted neutrino velocity is
\begin{equation}
    \hat{u}^\mu = \Lambda^\mu_{\phantom{\mu}\nu} \tilde{u}^\nu,
\end{equation}
where $\tilde{u}$ is the comoving velocity, $\hat{u}$ is the Eulerian velocity, and $\Lambda$ is the usual Lorentz boost matrix,
\begin{equation}
    \Lambda = 
    \begin{pmatrix}
        \gamma         & \gamma v_r                                 & \gamma v_\theta                               & \gamma v_\phi \\ 
        \gamma v_r      & 1 + (\gamma-1)\frac{v_r^2}{v^2}       & (\gamma-1)\frac{v_r v_\theta}{v^2}    & (\gamma-1)\frac{v_r v_\phi}{v^2} \\ 
        \gamma v_\theta & (\gamma-1)\frac{v_\theta v_r}{v^2} & 1 + (\gamma-1)\frac{v_\theta^2}{v^2}     & (\gamma-1)\frac{v_\theta v_\phi}{v^2} \\ 
        \gamma v_\phi   & (\gamma-1)\frac{v_\phi v_r}{v^2}    & (\gamma-1)\frac{v_\phi v_\theta}{v^2} & 1 + (\gamma-1)\frac{v_\phi^2}{v^2}
    \end{pmatrix}.
\end{equation}
Here, $\mathbf{v}$ is the velocity of the fluid with respect to the Eulerian frame, and $\gamma=(1-v^2)^{-1/2}$ is the Lorentz factor. When computing fluxes across all interfaces, we will encounter $\gamma v_r$, $\gamma v_\theta$, and $\gamma v_\phi$ terms 
again in the Lorentz boost. Physically, these terms are the fluid advection contributions to the neutrino transport. To ensure the stability of the scheme, these components are always handled using the upwind scheme as described in Section \ref{subsec:upwind}.

\subsubsection{Frame transformations} \label{subsubsec:frametransform}

The matrix $M^{\mu}_{\phantom{\mu}\nu}$ which transforms from the orthonormal Eulerian basis to the coordinate basis,
\begin{equation}
u^\mu = M^{\mu}_{\phantom{\mu}\nu} \hat{u}^\nu,
\end{equation}
needs to be chosen such that
$M^{\mu}_{\phantom{\mu}\alpha} g_{\mu\nu} M^{\nu}_{\phantom{\nu}\beta}=\eta_{\alpha\beta}$. For a general metric, this amounts to performing a Cholesky decomposition. Given a conformally flat metric, the transformation matrix and its inverse are given by
\begin{align}
  M
  &=\left(
  \begin{array}{cccc}
    \alpha^{-1} &  0 & 0 & 0 \\
    \alpha^{-1} \beta^x & \phi^{-2} & 0 & 0 \\
    \alpha^{-1} \beta^y & 0 & \phi^{-2} & 0 \\
    \alpha^{-1} \beta^z & 0 & 0 & \phi^{-2} \\
  \end{array}
  \right),
  \\
  M^{-1}
  &=\left(
  \begin{array}{cccc}
    \alpha & \phi^2 \beta^x & \phi^2 \beta^y & \phi^2 \beta^z \\
    0 & \phi^2 & 0 & 0 \\
    0 & 0 & \phi^2 & 0 \\
    0 & 0 & 0 & \phi^2 \\
  \end{array}
  \right).
\end{align}

\subsubsection{Ricci rotation of velocity components between cells} \label{subsubsec:ricci}

In order to properly handle geodesic transport of particles
between cells, we consider the equation of motion for the covariant
velocity components,
\begin{equation}
u_\mu = g_{\mu\nu} u^\nu,
\end{equation}
which can be obtained from the
super-Hamiltonian $H=g^{\mu\nu} u_\mu u_\nu$
with the covariant components $u_\mu$ serving as the canonical
``momenta''
\citep{1973grav.book.....M},
 as 
\begin{equation}
  \label{eq:eom}
\frac{\ud u_\mu}{\ud \lambda} =
- \frac{\pd g^{\alpha \beta}}{\pd x^\mu} u_\alpha u_\beta.
\end{equation}
Since the Hamiltonian does not depend on the
affine parameter (or canonical ``time'') $\lambda$,
its value is conserved along geodesics,
\begin{equation}
  \label{eq:hc}
\frac{\ud}{\ud \lambda} \left( g^{\alpha\beta} u_\alpha u_\beta \right) = 0.
\end{equation}
If the metric is taken to vary discretely from one cell to the next,
particles receive a
Hamiltonian ``kick'' $u_\mathrm{src} \rightarrow u_\mathrm{dst}$
when crossing a cell interface. The kick can be
calculated from 
the discrete version of Equation~(\ref{eq:hc}):
\begin{equation}
  \label{eq:dhc}
g^{\mu\nu}_\mathrm{dst} u_{\mu,\mathrm{dst}} u_{\nu,\mathrm{dst}}=g^{\mu\nu}_\mathrm{src} u_{\mu,\mathrm{src}} u_{\nu,\mathrm{src}}.
\end{equation}
Note that since the Hamiltonian varies only along one coordinate direction
across a cell interface, only the covariant velocity component
$u_i$
in the direction of the $i$-th coordinate will
receive a kick according to  Equation~(\ref{eq:eom}).

Now define $\delta g^{\mu\nu} = g^{\mu\nu}_{\mathrm{dst}} - g^{\mu\nu}_\mathrm{src}$ and
\begin{equation}
\delta u_{\mu} = \left\{
\begin{array}{c}
\delta, \quad \mu = i \\
0, \quad \mu \neq i
\end{array}
\right.,
\end{equation}
which allows us to rewrite Equation~(\ref{eq:dhc}) as
\begin{equation}
g^{\mu\nu}_{\mathrm{dst}} \delta u_\mu \delta u_\nu 
+ 2 g^{\mu\nu}_{\mathrm{dst}} u_{\mu,\mathrm{src}} \delta u_\nu
+ \delta g^{\mu\nu} u_{\mu,\mathrm{src}} u_{\nu,\mathrm{src}} = 0,
\end{equation}
and in terms of $\delta$ as
\begin{equation}
g^{ii}_{\mathrm{dst}} \delta^2
+ 2 g^{\mu i}_{\mathrm{dst}} u_{\mu,\mathrm{src}} \delta
+ \delta g^{\mu\nu} u_{\mu,\mathrm{src}} u_{\nu,\mathrm{src}} = 0.
\end{equation}
This is a quadratic equation for $\delta$, which formally has
two solutions,
\begin{equation}
\delta =
\frac{
-g^{\mu i}_\mathrm{dst} u_{\mu,\mathrm{src}}
\pm
\sqrt{\left( g^{\mu i}_\mathrm{dst} u_{\mu,\mathrm{src}}\right )^2
-g^{ii}_{\mathrm{dst}} \delta g^{\mu\nu} u_{\mu,\mathrm{src}} u_{\nu,\mathrm{src}}
}
}
{g^{ii}_{\mathrm{dst}}}.
\end{equation}
The plus solution is the correct solution for $u_{i,\mathrm{src}} > 0$, as it satisfies $\delta
\rightarrow 0$ for $\delta g^{\mu\nu} \rightarrow 0$. Similarly, the minus solution is the correct solution for $u_{i,\mathrm{src}} < 0$. A complication to this method arises if the kick necessary to conserve the Hamiltonian results in a velocity in the opposite direction through the interface, akin to the phenomenon of total internal reflection in refractive media. This manifests mathematically as the quadratic equation yielding no real solutions. We circumvent the issue by modifying the kick to become
\begin{equation}
    \delta = -u_{\mu,\mathrm{src}},
\end{equation}
and then scaling the other spatial components of $u$ so that energy is conserved. While this results in a small loss of momentum conservation, the error is insignificant in comparison to the error arising from the grid discretisation, and such cases only arise in the presence of extremely steep gradients in the metric. In summary, the kicked four-velocity is
\begin{equation}
    u_\mathrm{dst} = K(u_\mathrm{src}),
\end{equation}
where
\begin{equation}
    K(u_\mathrm{src})^\mu = g_\mathrm{dst}^{\mu\nu} (u_{\nu,\mathrm{src}}+\delta u_{\nu,\mathrm{src}}).
\end{equation}

\subsection{Computation of fluxes and update of the distribution function} \label{subsec:fluxes}
We express Equation (\ref{eq:boltzmann}) as the three-dimensional (unsplit) difference equation,
\begin{equation}
    \frac{f^{n+1} - f^{n}}{\Delta t} + \sum_{s\in\left\{a\pm\frac{1}{2},b\pm\frac{1}{2},c\pm\frac{1}{2}\right\}} \left[\mathbf{u}\cdot\mathbf{n}_s 
    (T_{s \rightarrow a,b,c}\left[\mathbf{f}_s^n\right])_{ijk}\ \frac{\Delta A_s}{\Delta V_{abc}}\right] = \mathfrak{C}\left[\mathbf{f}_{abc}\right]. \label{eq:fullstep}
\end{equation}
Here $\Delta t$ is the timestep and $\Delta V$ is the cell volume.
The advection term is summed over every interface $s$ of the cell, where $\Delta A_s$ is the interface area, $\mathbf{n}_s$ is the interface normal. The neutrino advection velocity
$\mathbf{u}$ is given by
the contravariant spatial components of the \emph{four-velocity}
at cell interfaces. The transformation operator $T$ is computed as explained in the 
previous sections; the subscript $s\rightarrow {abc}$
indicates that the transformation remaps the distribution
function $\mathbf{f}_s^n$ at the interface
 $s$ from interface frame to the comoving frame
in cell $(a,b,c)$ in real space.
For compactness and readability, we have omitted real
space and momentum space indices
for $f$, $\mathbf{u}$,
and the discretised collision integral $\mathfrak{C}$ (for which see Equation~\ref{eq:sourceterms_discrete} below).


Equation~(\ref{eq:fullstep}) gives
an updated distribution function in cell
$(a,b,c)$, but this updated distribution function
is still computed in the comoving frame
from the previous time step. After
accounting for updates to the distribution
function due to the fluxes at cell interfaces
and due to the collision integral, we therefore
still need to boost the distribution function
at time step $n+1$ from the comoving
frame at time step $n$ to the comoving
frame at time step $n+1$,
\begin{equation}
    \mathbf{f}^{n+1} \rightarrow
    T_{n\rightarrow n+1}
    [\mathbf{f}^{n+1}].
\end{equation}
This completes the transport time step.

In Sections \ref{subsec:lw}--\ref{subsubsec:smoothflux}, we will show how $f_s$ is obtained via either a Lax-Wendroff step, or an upwind step with linear reconstruction, both of which result in second-order convergence in space. For time integration, although the convergence of the advection terms is also second-order, the convergence of the collision term is only first-order due to the backward Euler differencing, which is usual for most transport codes. Defining the accuracy in angle and energy is non-trivial, given our particle-like treatment.

In optically thick regions, the source term is stiff, which would necessitate a restrictively small timestep in order to maintain stability in a fully explicit scheme. We circumvent this restriction by discretising the source term implictly in time (coupling the momentum space in a matrix inversion), while still solving the advection terms explicitly to avoid coupling in real space. For simplicity, we use a global timestep, limited by the Courant number in the smallest cell on the grid.

\subsubsection{Lax-Wendroff scheme}
\label{subsec:lw}
To recover the correct solution in the diffusion limit, we employ the Lax-Wendroff scheme, which provides second-order accuracy in both time and space. This amounts to replacing each $\mathbf{f}_s^n$ with $\mathbf{f}_s^{n+\hf}$, which is solved by advancing the solution by half a timestep in each spatial direction. For example, in the $a$ direction, this amounts to solving
\begin{multline}
    \frac{f_{a+\hf}^{n+\hf} - \frac{1}{2}\left(T_{a\rightarrow a+\hf}[\mathbf{f}_a^n] + T_{a+1\rightarrow a+\hf}[\mathbf{f}_{a+1}^n]\right)_{ijk}}{\frac{1}{2} \Delta t} \\+ \left(\mathbf{u}\cdot\mathbf{n}_{a+\hf}\right)\frac{\left(T_{a+1\rightarrow a+\hf}[\mathbf{f}_{a+1}^n] - T_{a\rightarrow a+\hf}[\mathbf{f}_a^n]\right)_{ijk}}{\Delta x} = \mathfrak{C}[\mathbf{f}_{a+\hf}^{n+\hf}]. \label{eq:halfstep}
\end{multline}
Here, $\Delta x = x_{a+1} - x_a$ is the distance between the cell centers. This equation is solved in the interface frame, so all transformations are to the interface frame. In the general multi-dimensional solution, this is repeated along each direction. We highlight that the $\frac{1}{2}\left(T_{a\rightarrow a+\hf}[\mathbf{f}_a^n] + T_{a+1\rightarrow a+\hf}[\mathbf{f}_{a+1}^n]\right)$ reconstruction term (which can alternatively be implemented as a geometric mean), together with the implicitly discretised source term on the right-hand side, provides the centered differencing crucial for reproducing the diffusion limit. Again, we have omitted extra real
space and momentum space indices
for $f$, $\mathbf{u}$,
and the collision integral $\mathfrak{C}$, implied to be $abc$ and $ijk$ except where specified. 

\subsubsection{Upwind scheme}
\label{subsec:upwind}
The spatially centered reconstruction of the Lax-Wendroff scheme produces an oscillatory solution in regions where the physical domain of dependence is one-sided, namely where the opacity is low and neutrinos stream freely. To maintain stability in these regions, we use an upwind scheme
\begin{equation}
    f_{a+\hf,{ijk}}^n = \left\{
        \begin{array}{cc} 
            \tilde{f}_{a,{ijk}}^n,     & \mathbf{u}_{a+\hf,ijk}\cdot\mathbf{n}_{a+\hf} \geq 0 \\
            \tilde{f}_{a+1,{ijk}}^n, & \mathbf{u}_{a+\hf,ijk}\cdot\mathbf{n}_{a+\hf} < 0
        \end{array}
    \right.,
\end{equation}
where $\tilde{f}_a^n$ and $\tilde{f}_{a+1}^n$ are linearly reconstructed values at $x_{a+\hf}$. Here, $\mathbf{u}$ is the velocity of the neutrino at the cell center expressed in the basis at the interface (i.e. $\mu=1$ is parallel with angular interfaces), to account for the angle of the cell interface relative to the cell center. Reconstruction is performed along characteristics, taking into account boosts along the path of the neutrino between cells, which recovers second-order accuracy in space. We apply the MC slope limiter \citep{1977JCoPh..23..276V} to suppress any remaining oscillations in the solution.

\subsubsection{Smooth flux switching}
\label{subsubsec:smoothflux}
To achieve a smooth transition between optically thick and thin regions, we interpolate between the interface distributions calculated by the Lax-Wendroff scheme and the upwind scheme using the formulation
\begin{equation}
    f_{a+\hf}^n = f_\mathrm{{a+\hf},LW}^{n+\hf} \left(1-w\right) + f_\mathrm{{a+\hf},upw}^n w,
\end{equation}
where the weights are motivated by the optical depth of the cell
\begin{equation}
    w = e^{-\left(\kappa_a + \kappa_s\right)\Delta x}.
\end{equation}
Since the neutrino opacity varies with energy, the flux switching must be performed separately for each energy cell.

\subsubsection{Boundary conditions} \label{subsec:boundaries}

In three dimensions, there exists on our grid formal boundaries at $r=0$, $r=r_\mathrm{max}$, $\theta=\theta_\mathrm{min}$, and $\theta=\theta_\mathrm{max}$. In practice, the only real boundary is at $r=r_\mathrm{max}$, where we apply an outflow boundary condition. The outflow boundary condition is appropriate for CCSN modelling where there are no external sources of neutrinos, because and neutrinos are travelling radially outward near the outer boundary. In two dimensions, we apply periodic boundary conditions in the $\phi$ direction, such that the domain is a wedge with angular size $\Delta\phi$, which we have the freedom to choose. In one dimension, we also apply periodic conditions to the $\theta$ such that the domain is a column, again with the freedom to choose the angular size $\Delta\theta$. In practice, we have found that choosing $\Delta\phi$ and $\Delta\theta$ so that all six interfaces of cells have a similar area produces the most accurate results without unnecessary restriction by the CFL criteria.

\subsection{Coupling to matter (source terms)} \label{subsec:sourceterms}
We model the production, absorption, and scattering of neutrinos by expressing the source term on the right-hand side of the Boltzmann equation (\ref{eq:boltzmann}) as
\begin{equation}
    \mathfrak{C}\left[f\right] = \ka \left( f_\mathrm{eq} - f \right) + \frac{\ks}{4\pi} \int_{4\pi}f\,\ud\Omega - \ks f, \label{eq:sourceterms}
\end{equation}
where $f_\mathrm{eq}$ is the equilibrium value of $f$, $\ka$ is the coefficient of absorption, and $\ks$ is the coefficient of scattering which we assume to be elastic and isotropic for the time being. The discretised form of this equation is
\begin{equation}
    \mathfrak{C}[f_{ijk}] = \kappa_{\mathrm{a},i} \left( f_{\mathrm{eq},i} - f_{ijk} \right) + \frac{\kappa_{\mathrm{s},i}}{4\pi} \sum_{\substack{j'=1,\dots,N_\mu\\k'=1,\dots,N_\Phi}} \left(f_{ijk}\,\Delta\Omega_{jk}\right) - \kappa_{\mathrm{s},i} f_{ijk}. \label{eq:sourceterms_discrete}
\end{equation}
Here, we have dropped the indices $a,b,c$ for compactness.

Since electron and antielectron neutrinos behave differently to the heavy-lepton species, and  present the greatest effect on the energetics of supernovae, we adopt the commonly used three-flavour approach, solving the transport equation independently for electron neutrinos, electron antineutrinos, and a third combined group of the remaining muon and tauon neutrinos and antineutrinos. The three-flavour approach provides a good balance between accuracy and computational efficiency, but our method can accomodate an arbitrary number of species. Eventually, with the implementation of weak magnetism terms, degeneracies between neutrinos and antineutrinos will be lifted, and more than three species will be required.

We use a subset of the neutrino interactions described by \cite{2002A&A...396..361R}. Using $f_\mathrm{eq}$, $\ka$, and $\ks$, we model absorption onto nucleons and nuclei, the Bremsstrahlung process in the one-particle approximation \citep{2015MNRAS.448.2141M,2015ApJS..219...24O}, and scattering on nucleons and heavi nuclei, for which the interaction rates are functions of the thermodynamic quantities of the fluid. We do not calculate non-isoenergetic scattering, which reduces the computational cost of implicitly solving the collision terms. While the set of interactions we use is incomplete, they are sufficient for the purposes of demonstrating the viability of our new scheme.

To model the backreaction of neutrinos on the fluid, we calculate hydrodynamical source terms for the lepton number, energy, and momentum
\begin{align}
    \label{eq:hydrosource1}
    Q_{Y_\mathrm{e}} &= 
    -\frac{1}{h^3}
    \int_{4\pi} \int_0^\infty \left( \mathfrak{C}_{\nu_e} - \mathfrak{C}_{\bar{\nu}_e} \right) \ud\epsilon\,\ud\Omega, \\
    Q_{E} &= - \frac{1}{h^3}
    \int_{4\pi} \int_0^\infty \left( \sum_{\nu_e, \bar{\nu}_e, \nu_{\mu/\tau}} \mathfrak{C} \right) \epsilon\,\ud\epsilon\,\ud\Omega, \\
    \mathbf{Q}_{M} &= - \frac{1}{h^3}\int_{4\pi} \int_0^\infty \left( \sum_{\nu_e, \bar{\nu}_e, \nu_{\mu/\tau}} \mathfrak{C} \right) \mathbf{p}\,\ud\epsilon\,\ud\Omega.
    \label{eq:hydrosource3}
\end{align}

The distribution function should always be non-negative ($f\geq 0$), and to obey Fermi statistics, should always be bounded ($f\leq 1$). The blocking factors in the opacity coefficients should in principle guarantee both these properties, but advection of neutrinos in momentum space can still cause cells to become overpopulated. This is unavoidable in a conservative finite-volume discretisation. In practice, we find that that the interpolation between Lax-Wendroff and upwind fluxes (Section \ref{subsubsec:smoothflux}) suppresses any non-monotonicities that might cause the distribution to become negative. However, the distribution can still overshoot beyond unity at sharp jumps in opacity (resulting in large fluxes across angular bins) or fluid velocity (resulting in large fluxes across energy bins). Naturally, this error diminishes with increasing resolution in momentum space, though this is not a computationally feasible solution. A naive correction for this error would result in a violation of conservation properties and lead to a secular drift in total number and/or energy.\footnote{Fixes have been devised to restore conservation at the expense of internal energy drift \citep{1993ApJ...405..669M,2004ApJS..150..263L}} We instead choose to maintain conservation properties and allow the distribution function to exceed unity. The only consequence is an error in the hydrodynamical source terms (Equations \ref{eq:hydrosource1}--\ref{eq:hydrosource3}). The opacity coefficients continue to be well-behaved and drive the distribution back towards $f\leq 1$, which causes the excess neutrinos to be absorbed back into the matter, where they may then be re-emitted into a higher energy bin. All in all, the transient overpopulation of lower energy bins results in a larger fraction of neutrinos interacting less strongly with the matter, which increases the trapping density.

\subsection{Implicit solution of the source terms}

\begin{figure*}
    \centering
    \includegraphics[width=0.49\linewidth]{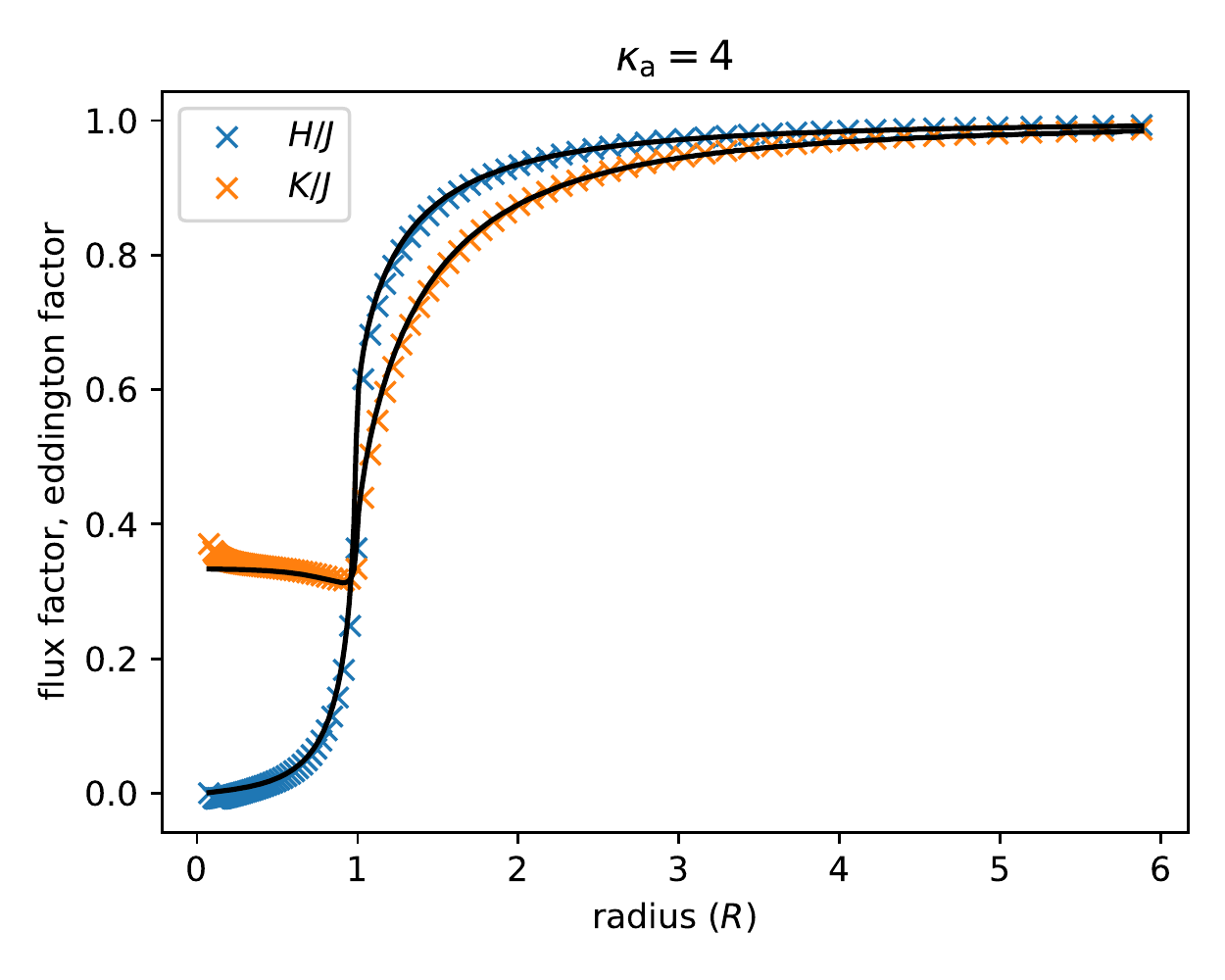}
    \hfill
    \includegraphics[width=0.49\linewidth]{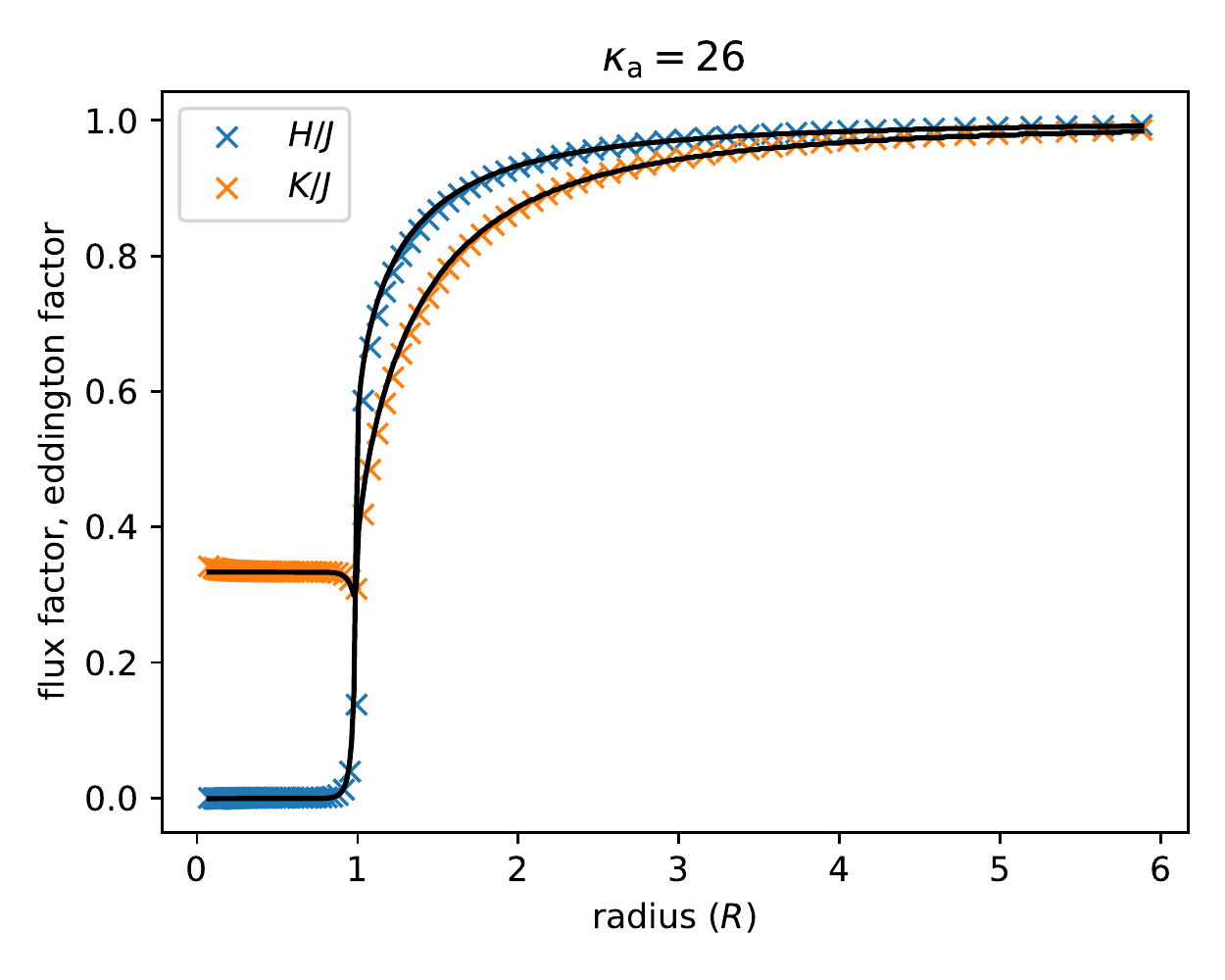}
    \caption{Radiating sphere test for $\tau=4$ representing ``low'' opacity (left) and $\tau=26$ representing ``high'' opacity. The flux factor and Eddington factor of the numerical solution (crosses) are compared against the exact solution (lines).}
    \label{fig:test_sphere}
\end{figure*}

In general, the solutions to Equations (\ref{eq:fullstep}) and (\ref{eq:halfstep}) can be found by solving a $N_\epsilon \times N_\mu \times N_\Phi$ linear system which couple together the $\epsilon$, $\mu$, and $\Phi$ dimensions. By approximating all scattering as elastic, this can be reduced to $N_\epsilon$ independent $N_\mu \times N_\Phi$ systems. We assume that scattering is both elastic and isotropic (\ref{eq:sourceterms}), for which there exists the computationally economical analytic solution
\begin{align}
    f_{jk}^{n+1} &= \frac{1}{1 + (\ka + \ks) \Delta t} 
    \times
    \Bigg[f_{jk}^{n} + \Delta t\, \ka f_\mathrm{eq} 
        \\
    &+ \frac{\ks \Delta t}{4\pi} \frac{\sum_{j'k'}\left(f_{j'k'}^{n}+\Delta t\, \ka f_\mathrm{eq} 
    + \Delta J_{j'k'}\right)\Delta\Omega}{1+\left(\ka+\ks\right)\Delta t} + \Delta J_{jk} \Bigg],
    \nonumber
\end{align}
where $\Delta J_{jk}$ is the advective flux divergence component of the difference equations (\ref{eq:fullstep}, \ref{eq:halfstep}), and the energy index $i$ has been omitted
for all terms.

\section{Code tests} \label{sec:codetests}

To demonstrate the quality of the transport scheme, we perform a suite of tests problems as described in \citet{2002A&A...396..361R} and \citet{2010ApJS..189..104M}, for which exact solutions can be obtained. We then demonstrate its capacity to be used in modelling a spherically symmetric core collapse supernova. Finally, we demonstrate that the scheme can handle two--dimensional transport problems with ease by performing a test in axisymmetry. 

\subsection{Classical radiating spheres} \label{subsec:radiatingsphere}

The radiating sphere test consists of a stationary homogeneous sphere $r \leq R$ of isotropically absorbing and emitting medium $f_\mathrm{eq}=\mathrm{const.}$ and $\ka=\mathrm{const.}$ within a vacuum $\ka=0$ for $r>R$. It is a commonly used benchmark which challenges the ability of radiative transfer methods to accurately model the transition from the diffusive to the free-streaming regime. The discontinuous jump in opacity poses a challenge even for state-of-the-art analytic two-moment closure methods, which have difficulty correctly reproducing the correct flux factor \citep{2017MNRAS.469.1725M}. It is a simple emulation of a supernova, where neutrinos escape the dense diffusive matter of the proto-neutron star though the discontinuity provides a more challenging transition compared to a real proto-neutron star surface. Nonetheless, the late-phase emission more closely resembles the homogeneous sphere test because the density gradient at the proto-neutron star surface becomes steeper. Demonstrating that a scheme is capable of reproducing the radiating sphere solution is the first step to proving its capacity to model CCSNe from core bounce through to the neutrino-driven wind phase. The analytic solution to this problem \citep[e.g.][]{1997A&A...325..203S}, as a function of radius $r$ and radiation angle $\mu$ is
\begin{equation}
    f(r,\mu) = f_\mathrm{eq} \left(1 - e^{-\ka s(r,\mu)}\right)
\end{equation}
where
\begin{equation}
    s(r,\mu) = \left\{ 
    \begin{array}{ll}
        r\mu + rg(r,\mu), & \qquad r<R, -1\leq\mu\leq1 \\
        2Rg(r,\mu),       & \qquad r\geq R, \sqrt{1-\left(\frac{R}{r}\right)^2}\leq\mu\leq 1 \\
        0, & \qquad \mathrm{else}
    \end{array}
    \right. ,
\end{equation}
and 
\begin{equation}
    g(r,\mu) = \sqrt{1 - \left(\frac{r}{R}\right)^2 \left(1 - \mu^2\right)}.
\end{equation}

\begin{table}
\centering
\caption{Mean square deviation 
\citep[Equation~(27) of][]{2017MNRAS.469.1725M} 
of the flux factor and Eddington factor between $R<r<2R$ for the radiating sphere test with $\tau=7500$. The first row (Boltzmann) shows the error of our scheme. The remaining rows show the error of commonly used closure relations from the results of \citet{2017MNRAS.469.1725M}.}
\label{tab:radiating_sphere_errors}
\begin{tabular}{lll} 
\hline
Method    & $\delta (H/J)$ & $\delta (K/J)$  \\ 
\hline
Boltzmann & 0.03           & 0.03  \\
Kershaw   & 0.13           & 0.32  \\
Wilson    & 0.05           & 0.14  \\
Levermore & 0.10           & 0.22  \\
ME        & 0.07           & 0.17  \\
MEFD      & 0.07           & 0.17  \\
Janka\_1  & 0.07           & 0.13  \\
Janka\_2  & 0.10           & 0.21  \\
\hline
\end{tabular}
\end{table}

\begin{figure*}
    \includegraphics[width=0.49\linewidth]{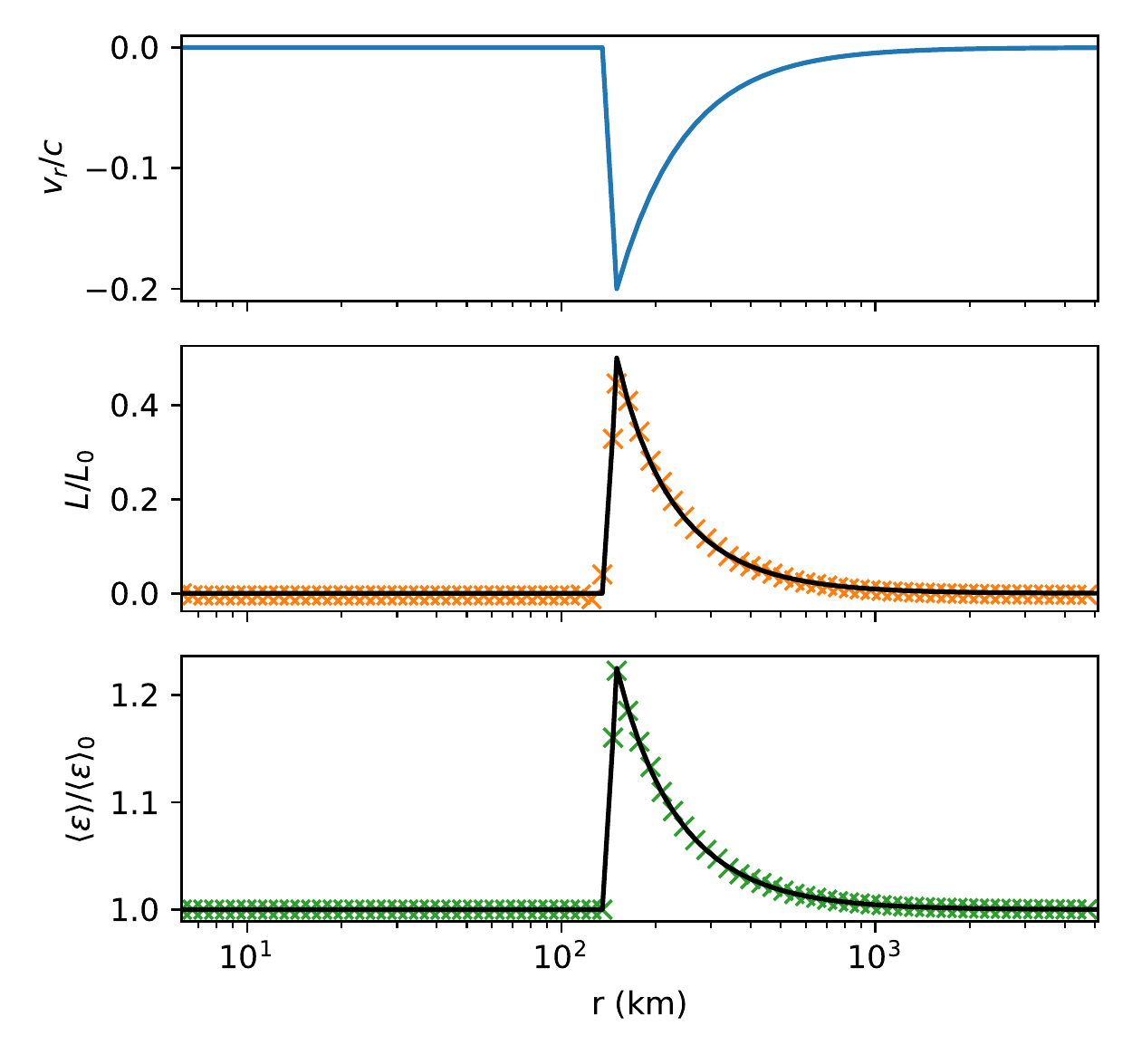}
    \hfill
    \includegraphics[width=0.49\linewidth]{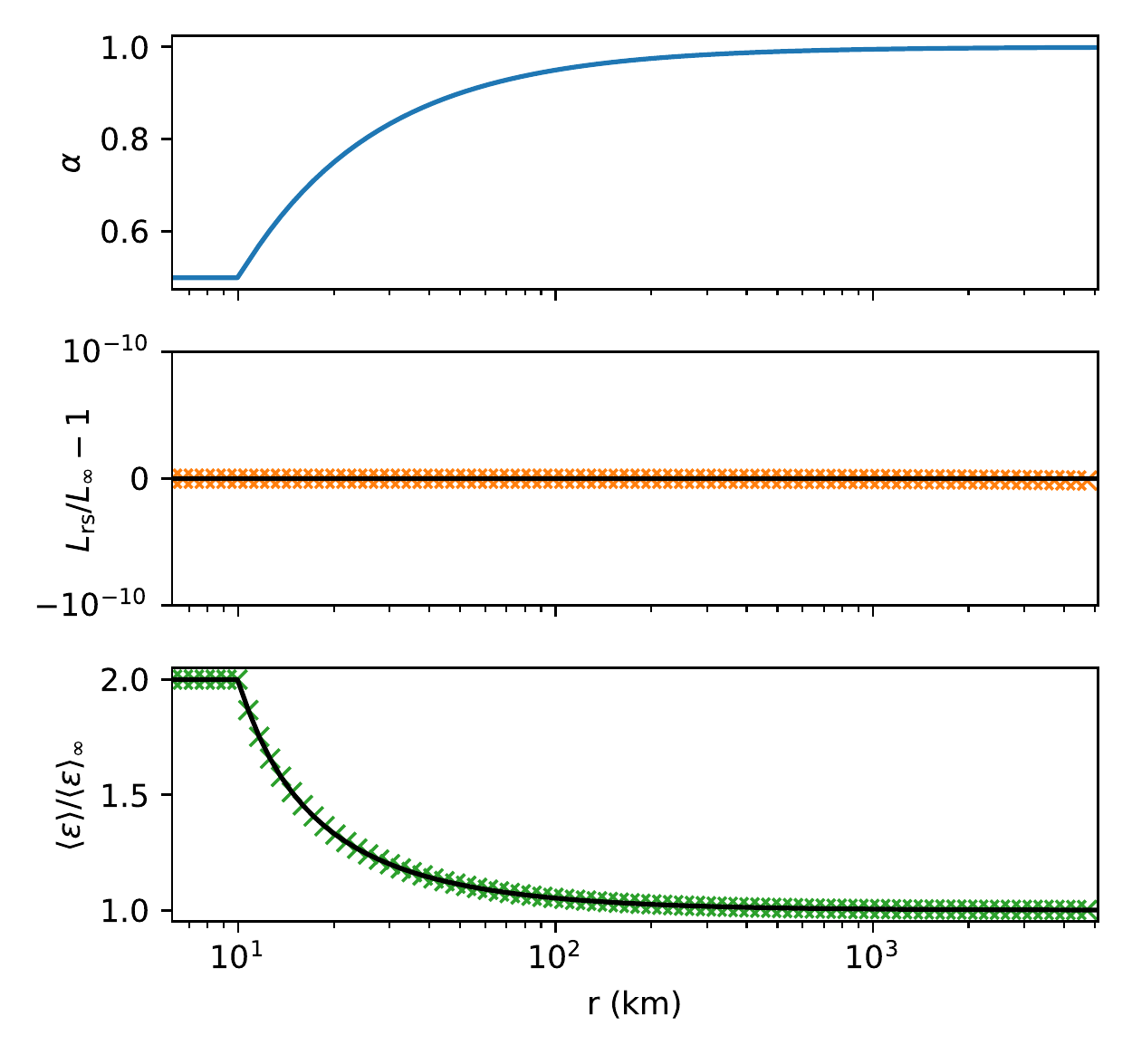}
    \caption{Setup and results of relativistic radiating sphere tests, repeating \citet{2010ApJS..189..104M}. Left: Radiating sphere with a shock velocity profile (top). We show the luminosity $L$ (middle) and mean energy $\langle\epsilon\rangle$ (bottom) as measured by a moving observer, normalised to the values $L_0$ and $\langle\epsilon\rangle_0$ just outside the radiating sphere, of our scheme (crosses) compared with the exact solution (lines). There is no redshift correction because there is no gravitational field. Right: Radiating sphere in a gravitational well (top). We show the redshift-corrected luminosity $L_\mathrm{rs}$ (middle) and mean energy $\langle\epsilon\rangle$ (bottom) as measured by a moving observer, normalised their the values at infinity $L_\infty$ and $\langle\epsilon\rangle_\infty$, of our scheme (crosses) compared to the exact solution (lines).}
    \label{fig:test_relativistic_sphere}
\end{figure*}

\begin{figure*}
    \includegraphics[width=\linewidth]{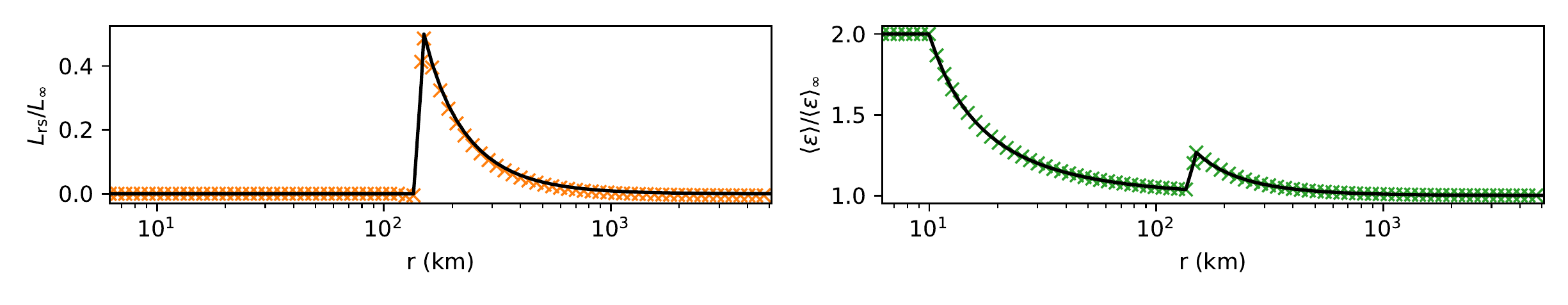}
    \caption{Radiating sphere, combining the shock velocity profile and gravitational well of the previous two tests. We show the redshift-corrected luminosity $L_\mathrm{rs}$ (left) and mean energy $\langle\epsilon\rangle$ (right) as measured by a moving observer, normalised their the values at infinity $L_\infty$ and $\langle\epsilon\rangle_\infty$, of our scheme (crosses) compared to the exact solution (lines).}
    \label{fig:test_relativistic_sphere_combined}
\end{figure*}

We repeat the radiating sphere tests performed by \cite{2002A&A...396..361R} and compare to the analytic solutions. We adopt a spatial resolution of $N_r=100$ logarithmically spaced
zones, an angular resolution of $N_\mu=9$ and $N_\Phi=4$, and a domain of $r_\mathrm{max} = 6R$. For the initial conditions $f=0$ everywhere. After a stationary state is reached, the flux factor $H/J$ and Eddington factor $K/J$ are compared with the exact solution. Here, $J,H,K$ are the zeroth, first, and second energy moments given by

\begin{align}
    J &= \frac{1}{4\pi} \int_{4\pi} f \epsilon \frac{\ud\Omega\,\epsilon^2\ud\epsilon}{h^3}, \\
    H &= \frac{1}{4\pi} \int_{4\pi} f \mu \epsilon \frac{\ud\Omega\,\epsilon^2\ud\epsilon}{h^3}, \\
    K &= \frac{1}{4\pi} \int_{4\pi} f \mu^2 \epsilon \frac{\ud\Omega\,\epsilon^2\ud\epsilon}{h^3}.
\end{align}

It is useful to perform this test at varying optical depths $\tau=\ka R$. We adopt the same two test parameters as \cite{2002A&A...396..361R}, one with low optical depth $\tau=4$ and one with high optical depth $\tau=26$. In both cases, we find excellent agreement with the analytic solution (\ref{fig:test_sphere}). Near the center, there is a slight deviation, an artefact of the grid singularity. The error is less pronounced at higher opacities. We also repeat the test by \cite{2017MNRAS.469.1725M} using $\tau=7500$. We compare the error of our scheme against the error of the closure schemes used in their test, and find that our scheme outperforms them all (Table \ref{tab:radiating_sphere_errors}).

\subsection{Relativistic radiating spheres}

We then proceed by repeating three tests from \cite{2010ApJS..189..104M}. We specify the test parameters in CGS units for ease of comparison to a real core-collapse and the existing literature \citep{2010ApJS..189..104M,2015ApJS..219...24O}. The first test verifies the scheme's ability to accurately account for Lorentz boosts. In this test (Figure \ref{fig:test_relativistic_sphere}, left), an optically thick radiating sphere is surrounded by a velocity profile similar to that of a stalled shock.  The radiating sphere has a radius of $4\,\mathrm{km}$, embedded in a velocity field
\begin{equation}
    v_r = \left\{ 
    \begin{array}{ll}
        0, & r < 135\,\mathrm{km} \\
        -0.2c\frac{r-135\,\mathrm{km}}{15\,\mathrm{km}}, & 135\,\mathrm{km}\leq r \leq 150\,\mathrm{km} \\
        -0.2c\left(\frac{150\,\mathrm{km}}{r}\right)^2, & 150\,\mathrm{km} \leq r
    \end{array}
    \right. .
\end{equation}

The second test (Figure \ref{fig:test_relativistic_sphere}, right) verifies the scheme's ability to accurately account for gravitational redshift in the presence of a curved spacetime. In this test, the same optically thick radiating sphere is embedded into a gravitational potential
\begin{equation}
    \alpha = \left\{ 
    \begin{array}{ll}
        0.5, & r < 10\,\mathrm{km} \\
        0.5 + 0.5\times\left(1-\frac{10\,\mathrm{km}}{r}\right), & r\geq 10\,\mathrm{km} 
    \end{array}
    \right. .
\end{equation}
The central value of the lapse function is $\alpha = 0.5$ and gradient $\frac{\ud \alpha}{\ud r}$ is more extreme than in a typical core-collapse supernova, and thus provides a good demonstration of the robustness of our scheme. In particular, we deliberately include a discontinuity in $\frac{\ud \alpha}{\ud r}$ at $10,\mathrm{km}$ to probe for any stability issues. We set the conformal factor to $\phi = \alpha^{-1/2}$.

The third test (Figure \ref{fig:test_relativistic_sphere_combined}) is a combination of the first two. To demonstrate our scheme's ability to handle even the most difficult scenarios, we choose $\ka = 100$ to obtain an optically thick sphere in \emph{only} the highest energy cell, and $\ka = 0$ everywhere else, resulting in a discontinuity in energy space. We use $N_\epsilon = 12$ energy cells ranging from $4\,\mathrm{MeV}$ to $240\,\mathrm{MeV}$ for a modest energy resolution of $\Delta\epsilon/\epsilon \approx 0.36$. 

We test for the accurate advection of neutrinos in energy space by comparing the redshift-corrected luminosity
\begin{equation}
    L_\mathrm{rs} = 16\pi\alpha^2 \phi^4 H r^2
\end{equation} and the mean energy $\langle\epsilon\rangle = H/\mathcal{H}$ as measured in the comoving observer frame with the analytic solutions. Here, 
\begin{equation}
    \mathcal{H} = \frac{1}{4\pi} \int_{4\pi} f \mu\frac{\ud\Omega\,\epsilon^2\ud\epsilon}{h^3}
\end{equation}
is the neutrino number density. In the vacuum region outside the sphere, the collision integral vanishes, and $L_\mathrm{rs}$ and $\langle\epsilon\rangle$ as measured by a moving observer obey \citep{2010ApJS..189..104M}
\begin{equation}
    \frac{1+v_r}{1-v_r}L_\mathrm{rs} = \mathrm{const}, \qquad
    \gamma \left(1+v_r\right) \alpha \langle\epsilon\rangle = \mathrm{const},
\end{equation}
where $v_r$ is the observer velocity, which we choose to be the same as the fluid velocity, and $\gamma$ is the Lorentz factor. In the second test where $v_r=0$, $L_\mathrm{rs}$ should be constant, which our scheme reproduces excellently.

\begin{figure*}
    \centering
    \includegraphics[scale=0.7]{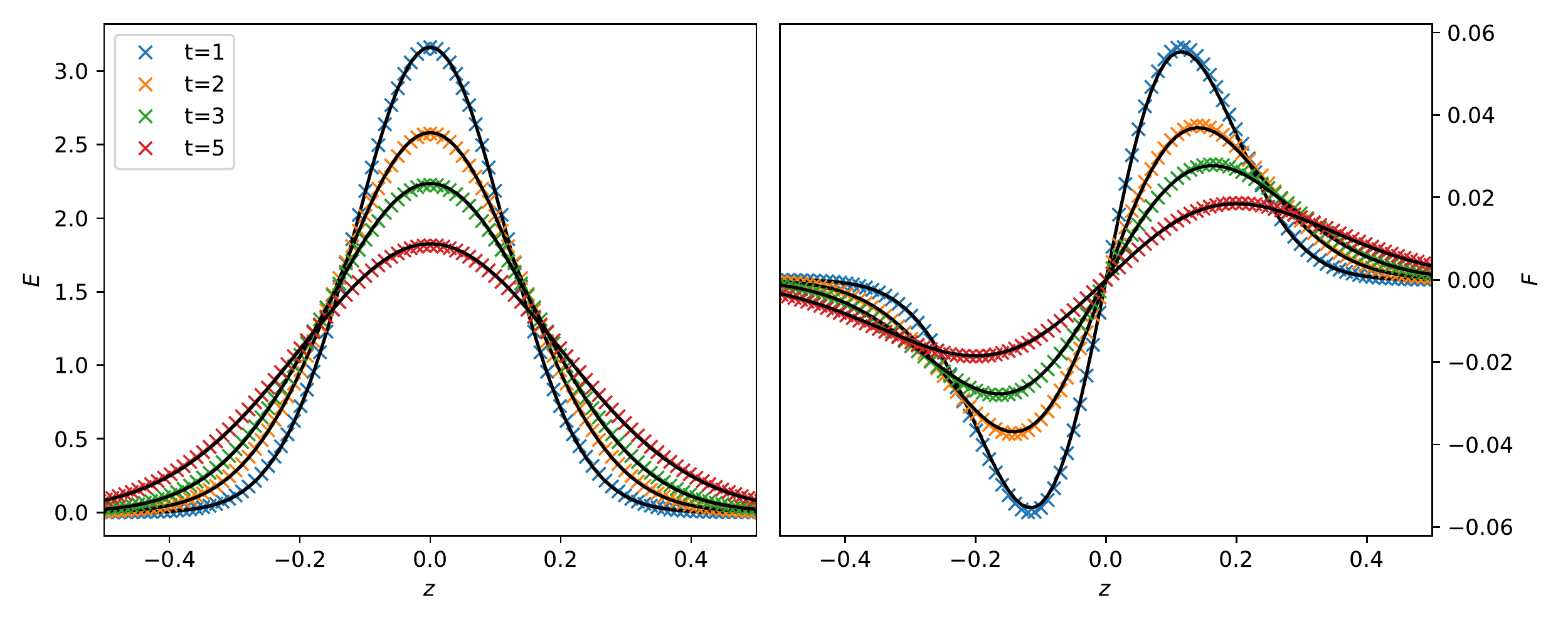}
    \caption{Diffusion wave test in planar geometry. The initial conditions are given at $t=1$. The numerical solution is sampled at $t=2,3,5$ (crosses), and compared with the exact solution (lines).}
    \label{fig:diff_wave_planar}
\end{figure*}

\begin{figure*}
    \includegraphics[width=0.49\linewidth]{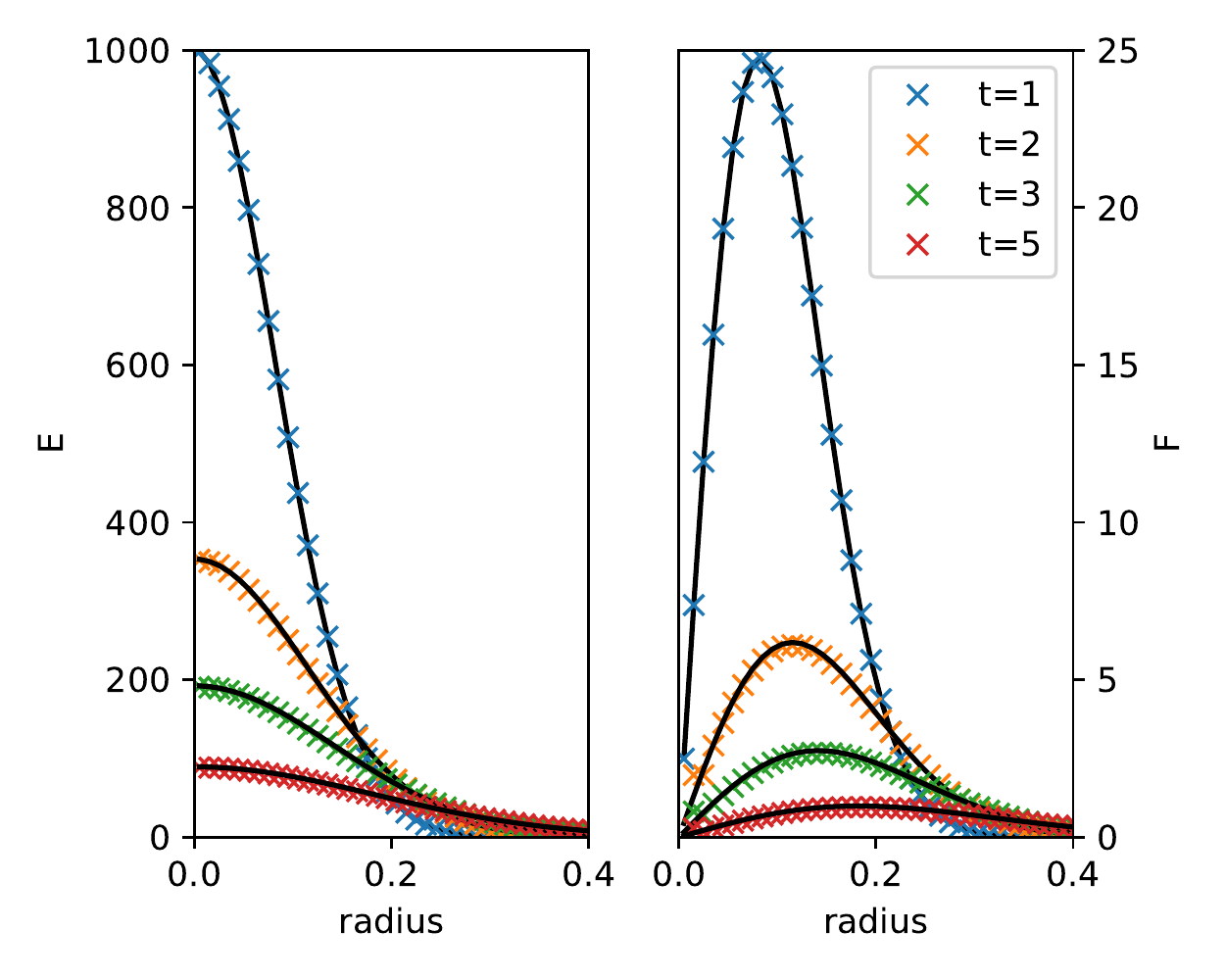}
    \hfill
    \includegraphics[width=0.49\linewidth]{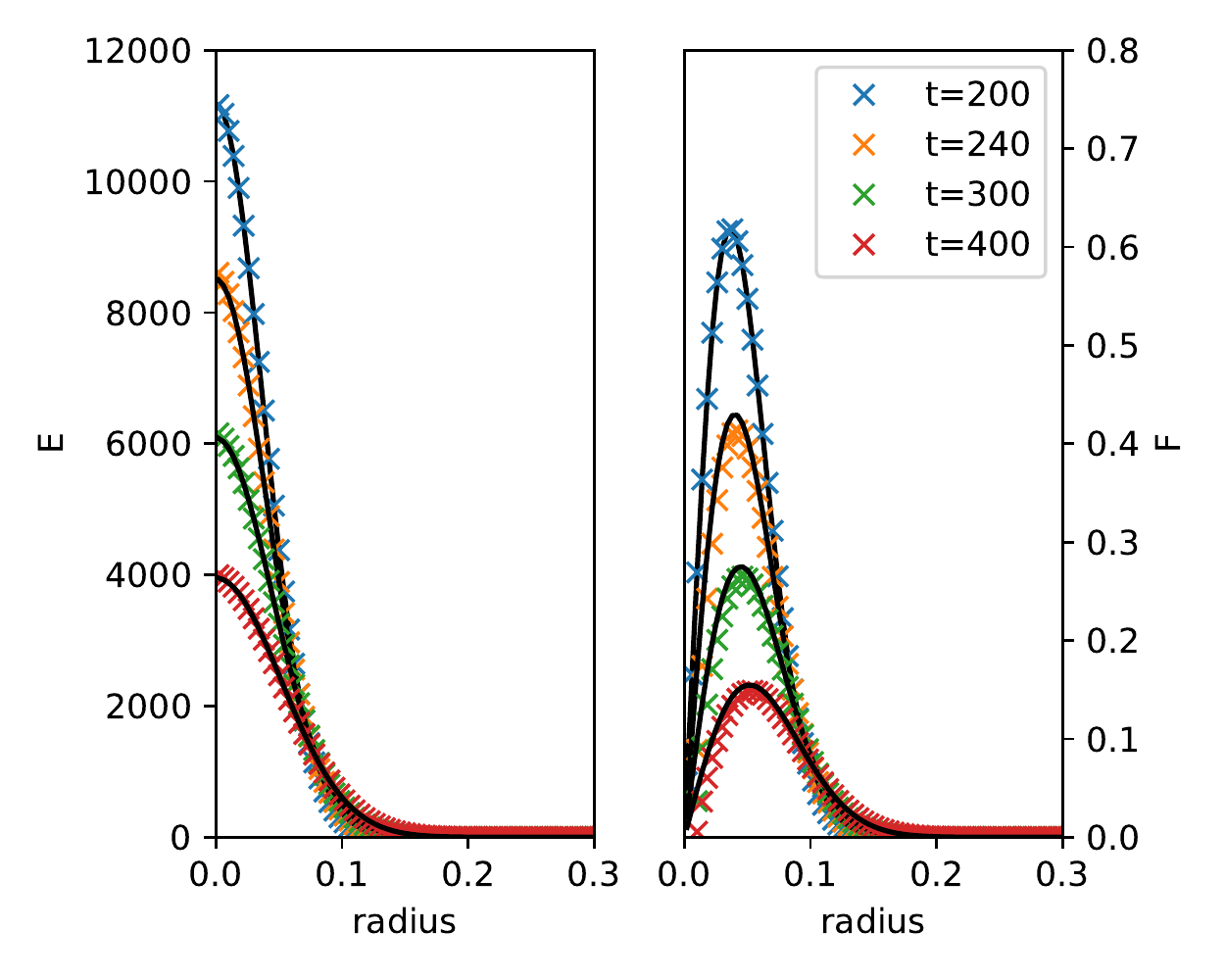}
    \caption{Diffusion wave test in spherical polar geometry. comparing the numerical solutions (crosses) to the exact solutions (lines). Left: Energy density and flux for $\ks=10^2$ ($\mathrm{Pe}=1$), initialised at $t=1$ and sampled at $t=2,4,5$. Right: Energy density and flux for $\ks=10^5$ ($\mathrm{Pe}=1000$), initialised at $t=200$ and sampled at $t=240,300,400$.}
    \label{fig:diff_wave_spherical}
\end{figure*}

\subsection{Diffusion wave}
To show that our scheme is able to reproduce the correct fluxes in the diffusion limit, we test the diffusion of a Gaussian distribution in an optically thick medium. First, we test our scheme in planar geometry, which allows us to examine the Lax-Wendroff fluxes in isolation, without the rotation terms associated with the spherical polar grid. The exact solution for the evolution of a Dirac delta function in the diffusion limit is
\begin{align}
    E &= \sqrt{\frac{\ks}{t}}\exp\left(\frac{-3\ks z^2}{4t}\right), \\
    F &= \frac{z}{2t}E.
\end{align}
Here, $E=4\pi J$ is the energy density, and $F=4\pi H$ is the energy flux. Our grid spacing is $\Delta z=0.01$, and we set $\ks=100$ so that $\mathrm{Pe}=1$. The initial conditions are set as the exact solution at $t=1$. We sample our numerical solution at $t=2,3,5$, and find excellent agreement with the exact solution (Figure \ref{fig:diff_wave_planar}). Next, we perform the diffusion wave test on a spherical grid to incorporate the effect of geometric rotations. We use an identical setup to \cite{2000MNRAS.317..550P}, with $\Delta r = 0.01$, repeated for two opacities $\ks=10^2$ ($\mathrm{Pe}=1$) and $\ks=10^5$ ($\mathrm{Pe}=1000$). The exact solution in the diffusion limit is
\begin{align}
    E &= \left(\frac{\ks}{t}\right)^{3/2}\exp\left(\frac{-3\ks r^2}{4t}\right), \\
    F &= \frac{r}{2t}E.
\end{align}
We again set the initial conditions as the exact solution at $t=1$ for the $\ks=10^2$ case and $t=200$ for the $\ks=10^5$ case. We find acceptable agreement with the exact solution in both cases (Figure \ref{fig:diff_wave_spherical}). The small error near the grid singularity $r=0$ is due to the interpolation between angle groups for lateral fluxes.

\begin{figure*}
    \centering
    \includegraphics[scale=0.7]{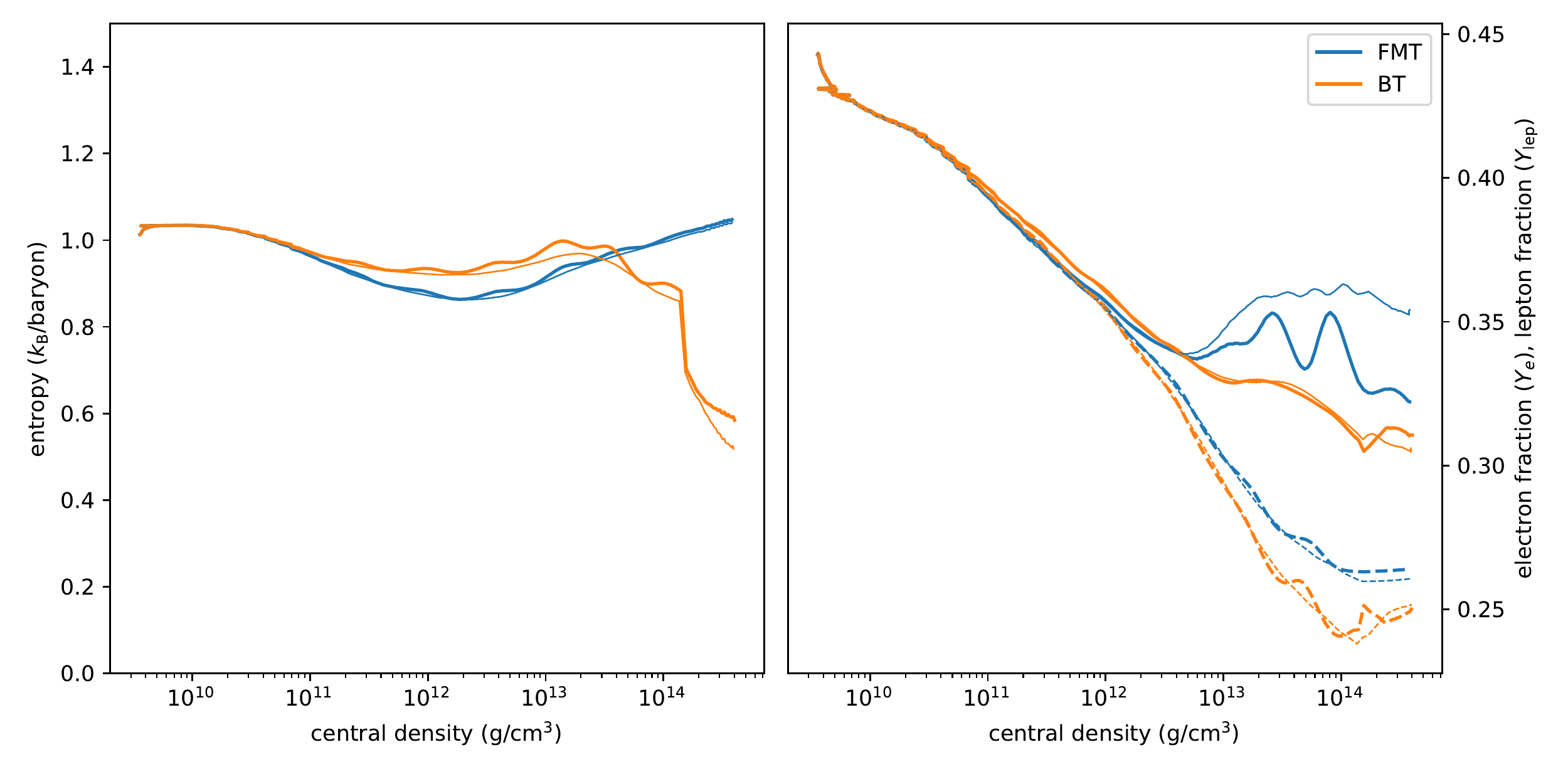}
    \caption{Central entropy (left), total lepton fraction
    $Y_\mathrm{lep}$ (right, solid lines), and electron fraction $Y_\mathrm{e}$ (right, dashed lines) as a function of central density during collapse. Blue lines show the \textsc{FMT} results, and orange lines show the results from our new scheme. The runs with $N_\epsilon=12$ (thick lines) are compared with the runs with  $N_\epsilon=24$ (thin lines) to show that the cause of the oscillations at $\rho_\mathrm{c}>10^{12}\,\mathrm{g}/\mathrm{cm}^3$ are caused by discretisation in energy space.}
    \label{fig:s20_central}
\end{figure*}

\begin{figure*}
    \includegraphics[width=\linewidth]{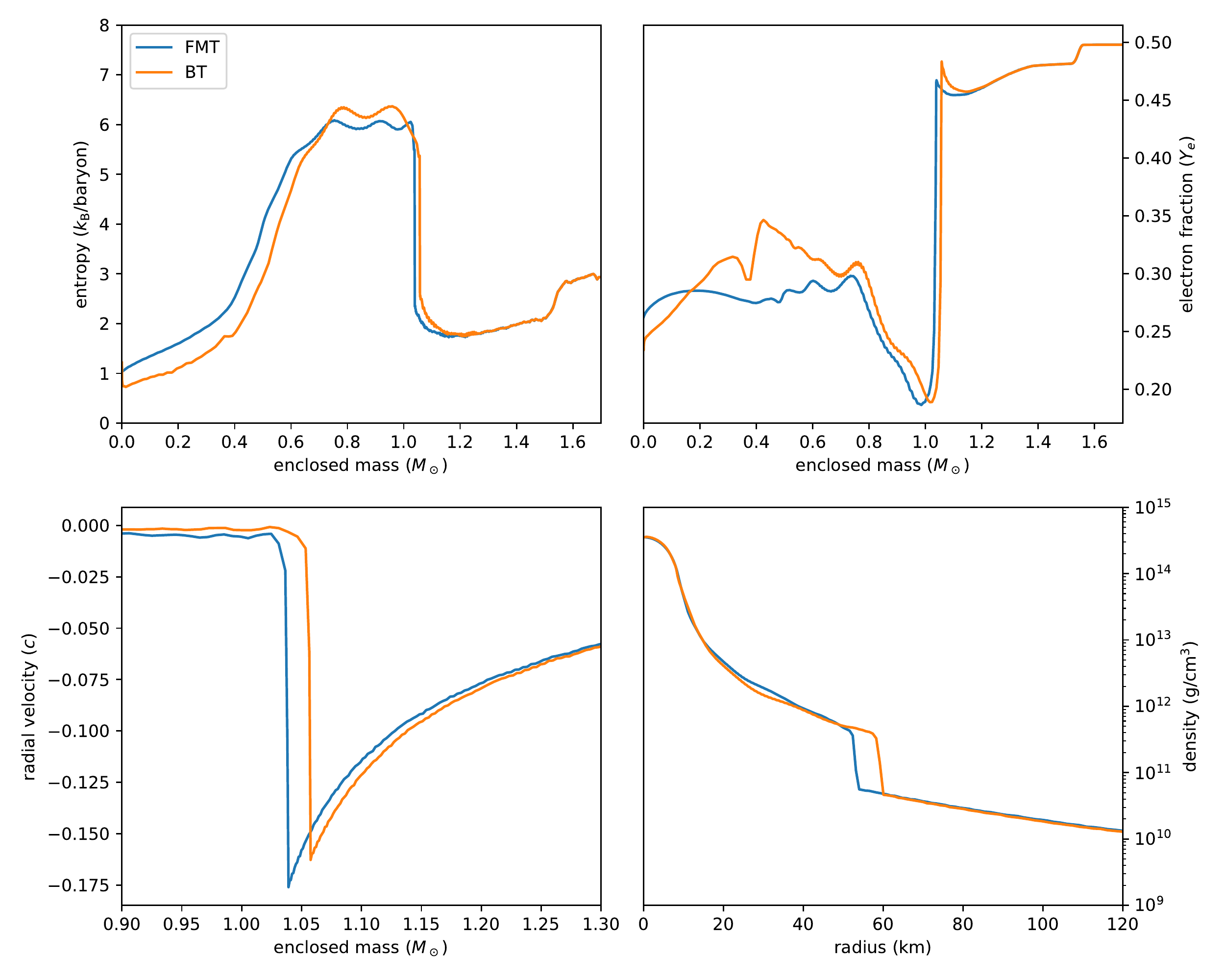}
    \caption{Profiles of the hydrodynamical quantities at $3\,\mathrm{ms}$ after bounce. We show entropy as a function of enclosed mass (top left), electron fraction as a function of enclosed mass (top right), radial velocity as a function of enclosed mass (bottom left), and density as a function of radius (bottom right). Blue lines show the \textsc{FMT} results, and orange lines show the results from our new scheme.}
    \label{fig:s20_pb_hydro}
\end{figure*}

\begin{figure*}
    \includegraphics[width=\linewidth]{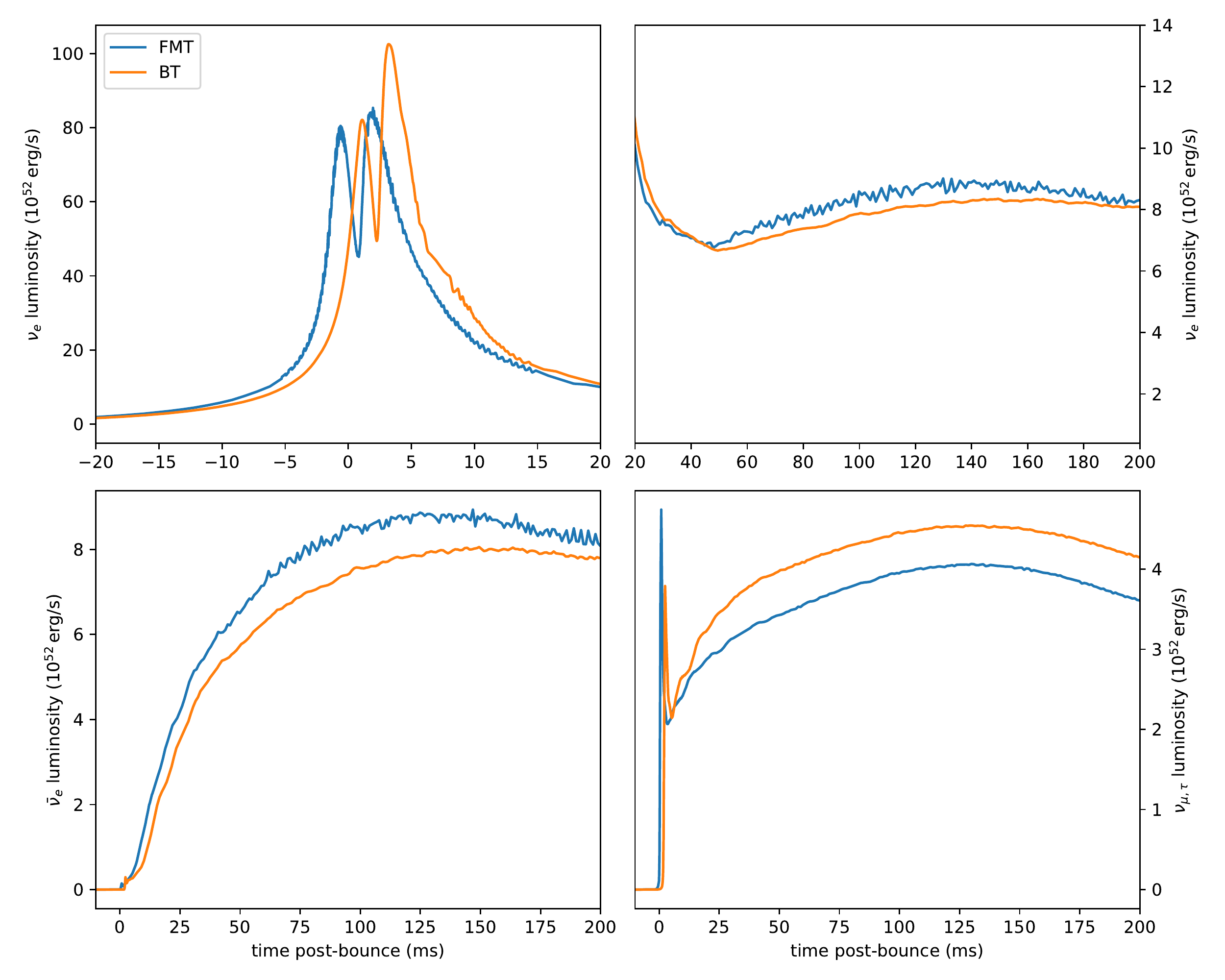}
    \caption{Neutrino luminosities for each flavor group as measured by an observer at $500\,\mathrm{km}$ as a function of time after bounce. We show the electron neutrino luminosity in the upper panels, with the upper left panel focusing on the neutronization burst. We show the electron antineutrion and heavy lepton neutrino luminosities in the lower left and lower right panels. Blue lines show the \textsc{FMT} results, and orange lines show the results from our new scheme.}
    \label{fig:s20_luminosity}
\end{figure*}

\begin{figure*}
    \includegraphics[width=\linewidth]{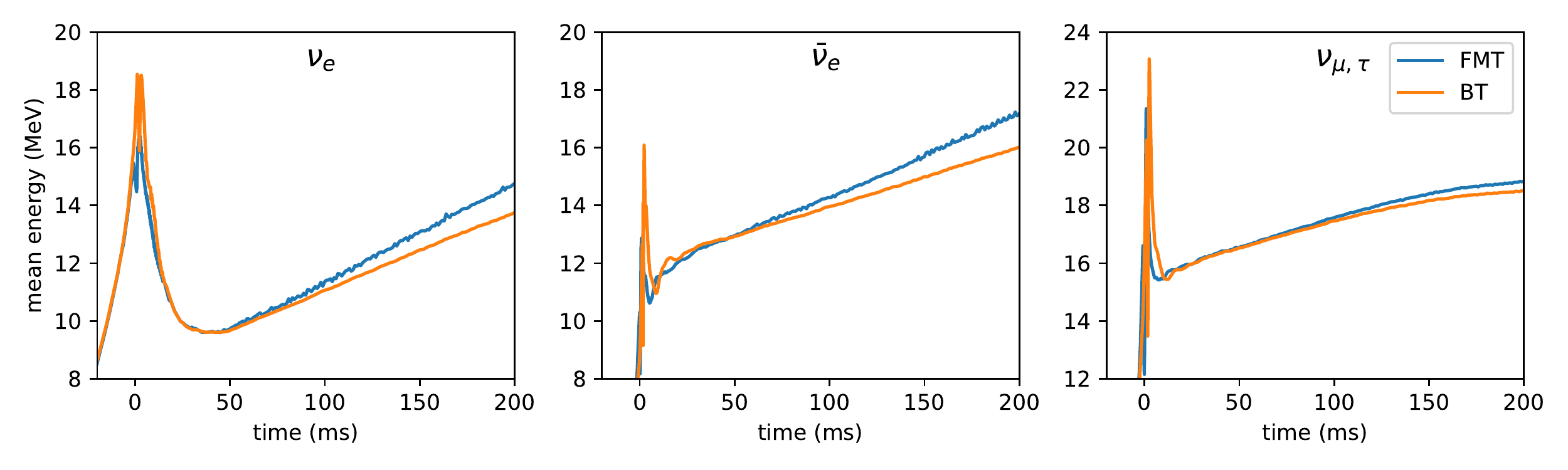}
    \caption{Mean energy of electron neutrinos, electron antineutrinos, and heavy lepton neutrinos (from left to right) as measured by an observer at $500\,\mathrm{km}$ as a function of time after bounce. Blue lines show the \textsc{FMT} results, and orange lines show the results from our new scheme.}
    \label{fig:s20_energy}
\end{figure*}

\subsection{Core collapse in spherical symmetry} \label{subsec:cctest}
Having demonstrated that our scheme can handle these rigorous test problems, we move on to a spherically symmetric core-collapse simulation in general relativity as a test for time-dependent effects when coupled to a hydrodynamics code. We couple our Boltzmann scheme to the general relativistic hydrodynamics code \textsc{CoCoNuT}, and evolve a $20\Msun$ model from collapse up to over $200\,\mathrm{ms}$ after bounce. 
The purpose of this simulation is to demonstrate that the coupling 
is stable and reproduces the familiar dynamics of the collapse
and accretion phase. The test is not meant to rigorously quantify the
accuracy of the code in dealing with the full supernova problem, which
would be premature for the first implementation of a completely new
numerical method while upgrades are still ongoing. For the
purpose at hand, it is therefore appropriate to use simplified opacities
and use a relatively simple, but reasonably accurate scheme
for a reference solution. Here we compare our
results to the \textsc{FMT} scheme, which has been described
and tested against the variable Eddington factor code
\textsc{Vertex} \citep{2002A&A...396..361R,2010ApJS..189..104M} by \citep{2015MNRAS.448.2141M}.
A more thorough validation
against other 1D transport codes by
a more detailed code
comparison as in \citet{2005ApJ...620..840L,2010ApJS..189..104M,2015ApJS..219...24O,2015MNRAS.453.3386J,2018JPhG...45j4001O} is not expedient as long as critical microphysics, i.e.\ neutrino-electron scattering,  during the collapse phase. In the present simulation,
we switch off neutrino scattering off heavy nuclei to mimic the final low-$Y_\mathrm{e}$ conditions in the core, but nonetheless the collapse dynamics is too different from models with  more sophisticated microphysics to allow a meaningful comparison.

We adopt the same grid coordinates for the neutrino transport as the hydrodynamics. We couple the transport and
the hydrodynamics solvers using an operator-split method, where the fluid quantities are first updated, and then used to compute the neutrino source terms for
the next timestep. The CFL limit on the timestep for neutrino transport is smaller than that on the fluid, so we perform multiple substeps to match the hydrodynamic timestep, after which the neutrino distribution is used to compute the hydrodynamic source terms for the subsequent step. We compare the results directly with a simulation performed using the \textsc{FMT} scheme, keeping the initial conditions, the hydrodynamical scheme, and microphysical inputs identical, so that we can isolate the differences arising from our new transport scheme. In the following, we refer to the two runs as \textsc{BT} 
(Boltzmann transport)
and \textsc{FMT}. In both schemes, we use $N_\epsilon = 12$ energy cells ranging from $4\,\mathrm{MeV}$ to $240\,\mathrm{MeV}$. In \textsc{BT}, we use an angular resolution of $N_\mu=9$ and $N_\Phi=4$.

In Figure~\ref{fig:s20_central}, we show the evolution of the entropy, electron number fraction $Y_\mathrm{e}$, and total lepton number fraction $Y_\mathrm{lep}=Y_\mathrm{e}+Y_{\nu_e}-Y_{\bar{\nu}_e}$ in the core as a function of central density. During collapse, the core density $\rho_\mathrm{c}$ rapidly rises from $\mathord{\sim} 10^9\,\mathrm{g}/\mathrm{cm}^3$ to $\mathord{\sim} 10^{14}\,\mathrm{g}/\mathrm{cm}^3$. Neutrinos are released, corresponding to the decline in electron number. At densities below $\mathord{\sim} 10^{12}\,\mathrm{g}/\mathrm{cm}^3$, matter is transparent to neutrinos, allowing neutrinos escape and carry away lepton number. 
Since scattering off nuclei is turned off, trapping
is delayed to densities considerably 
 above $\mathord{\sim} 10^{12}\,\mathrm{g}/\mathrm{cm}^3$.
Trapping only sets in around 
$\mathord{\sim} 10^{13}\,\mathrm{g}/\mathrm{cm}^3$
when $Y_\mathrm{e}$ and  $Y_\mathrm{lep}$ start to diverge.
We note that the electron fraction in the two runs follows
a slightly different trajectory. 
This is because the \textsc{FMT} scheme does not advect trapped neutrinos
but assumes instantaneous adjustment to a stationary transport
solution; hence the neutrino phase-space blocking terms and 
the source term $Q_{Y_\mathrm{e}}$ for the electron fraction
no longer agree in the trapping regime.
After trapping the central lepton fraction should ideally 
remain constant if we used more accurate opacities, and after equilibrium between the matter 
and the neutrinos is established, the entropy should
remain constant as well. However, we still find non-negligible changes in the lepton fraction and entropy
beyond densities of $10^{13}\,\mathrm{g}/\mathrm{cm}^3$
in the \textsc{BT} run. This is likely due to a combination
of factors. The fact that the \textsc{FMT} run
(i.e., with a scheme that reproduces trapping if
scattering off nuclei is included)
shows a decrease of $Y_e$ up to a density of 
$10^{14}\,\mathrm{g}/\mathrm{cm}^3$ suggests that without
the scattering off nuclei as the major source of scattering
opacity trapping simply remains incomplete until
very high densities. This suggests that part
of the problem lies in our choice of opacities.
However, there are also indications
of a numerical accuracy problem since we also note
non-monotonic changes in lepton fraction and entropy
above densities of $10^{14}\,\mathrm{g}/\mathrm{cm}^3$
where trapping should be complete even without coherent
scattering on nuclei. 
The artifact seems
insensitive to the energy resolution; doubling the
number of energy groups did not improve entropy
and lepton fraction conservation in the trapping regime.\footnote{Note that 
entropy, lepton fraction, and electron fraction are also oscillatory in
the \textsc{FMT} run from $\rho_\mathrm{c}\sim 10^{13}\texttt{-}10^{14}\,\mathrm{g}/\mathrm{cm}^3$. This \emph{is} due to the low energy resolution. As discussed by \cite{1993ApJ...405..637M}, these oscillations are caused by the  finite resolution of the Fermi distribution, and become more pronounced at higher densities because the higher energy bins account for a large fraction of the phase space. Repeating the simulation using twice as many energy bins ($N_\epsilon = 24$) increases the frequency and decreases the amplitude of these oscillations, confirming that this is indeed the cause.}
Closer inspection of the \textsc{BT} run reveals that
a noticeable blip at the centre in the entropy and electron fraction profiles, because the neutrino flux factor does
not perfectly asymptote to zero at the origin. This numerical artifact affects a few of the innermost
zones, and is rather sensitive to numerical details,
such as the precise choice of boundary conditions.
However, the blip eventually stabilises, and there is
no strong long-term drift in the central entropy
and electron fraction that would compromise stability
during the post-bounce phase. 
The treatment of the
innermost zones will be further optimised in future
runs with more complete opacities.

After bounce, 
\textsc{BT} produces the expected
behaviour in the hydrodynamics and the neutrino emission.
We adopt the commonly used definition of bounce as the time when the core entropy first exceeds $3\,k_\mathrm{B}/\mathrm{baryon}$. The mass enclosed within the radius of shock formation is $0.52\,\Msun$ in \textsc{BT} and $0.44\,\Msun$ in \textsc{FMT}.
In Figure (\ref{fig:s20_pb_hydro}), we show radial profiles of the hydrodynamic quantities $3\,\mathrm{ms}$ after bounce. Aside from an insignificant difference in the position of the shock, the two runs agree reasonably well.
\textsc{BT} is able to stably follow the propagation of
the shock from the optically thick to the optically
thin regime. 

In Figure (\ref{fig:s20_luminosity}), we show the neutrino luminosities as measured by an observer at a radius of $500\,\mathrm{km}$. The peak in \textsc{BT} is delayed by $2\,\mathrm{ms}$ because the burst takes a small
amount of time to reach $500\,\mathrm{km}$, whereas
the luminosity at the observer radius instantly
increases once the post-shock matter becomes optically
thin in \textsc{FMT} because a stationary transport
solution is used. The electron neutrino luminosity in \textsc{BT} peaks at a higher value than \textsc{FMT} during the neutronization burst, but agrees well afterwards. One
should note the rather unusual shape of the burst
with a high peak luminosity and
 two humps of roughly equal height, which is purely
due to our choice of opacities. During the accretion phase, the antineutrino luminosity is lower, whereas the heavy-lepton neutrino luminosity is higher.

In Figure (\ref{fig:s20_energy}), we show the mean neutrino energy as measured by an observer at a radius of $500\,\mathrm{km}$. In \textsc{BT}, the mean energy peaks at roughly $2\,\mathrm{MeV}$ higher during the neutronization burst, but agree reasonably well afterwards. At later times, the mean energy of all three species in \textsc{BT} is lower.
During the early post-bounce phase the differences in
the luminosities and mean energies between \textsc{BT}
and \textsc{FMT} are below $10\%$, and give a
quantitative indication of the errors in the \textsc{FMT} scheme
that stem from the transport approximation (but not
from the microphysics).

Although neither run accurately maintains a constant core entropy and lepton fraction during bounce, this test sufficiently demonstrates the basic viability of our algorithm. We defer a comparison with other state-of-the-art models run using other Boltzmann transport solvers until our collision integral has been updated to include the most recent microphysics.

Both runs were performed on the same supercomputer (OzSTAR), providing us a means to estimate the relative computational cost. \textsc{BT} was run using $24$ cores, while \textsc{FMT} was run using 1 core. The elapsed wall clock time per time step was 28 times longer in \textsc{BT}, which translates to 672 times more core-hours consumed. It is unsurprising that \textsc{FMT} is much faster, given that it is a far more approximate scheme. We emphasize, that there is a great amount of effort still need to optimize and parallelize the code before it is affordable to run in full 6D, and these timings are only rough estimates. On one hand, more expensive opacities and parallelization overheads will further increase computational cost, but on the other hand, our implementation of the algorithm is currently far from optimized.

\subsection{Tests in axisymmetry} \label{subsec:2dtests}
Up until now, we have performed tests in spherical symmetry, but the real advantage of our method is that it can be easily extended to two and three dimensions. We limit our tests to stationary problems in axisymmetry, which can still be computed using modest computational resources. For axisymmetric problems, periodic boundary conditions are applied in the $\phi$-direction, thus we expect results to be identical on a grid in 3D with the same $\Delta\phi$.

\subsubsection{Radiating sphere}
\begin{figure*}
    \includegraphics[width=0.49\linewidth]{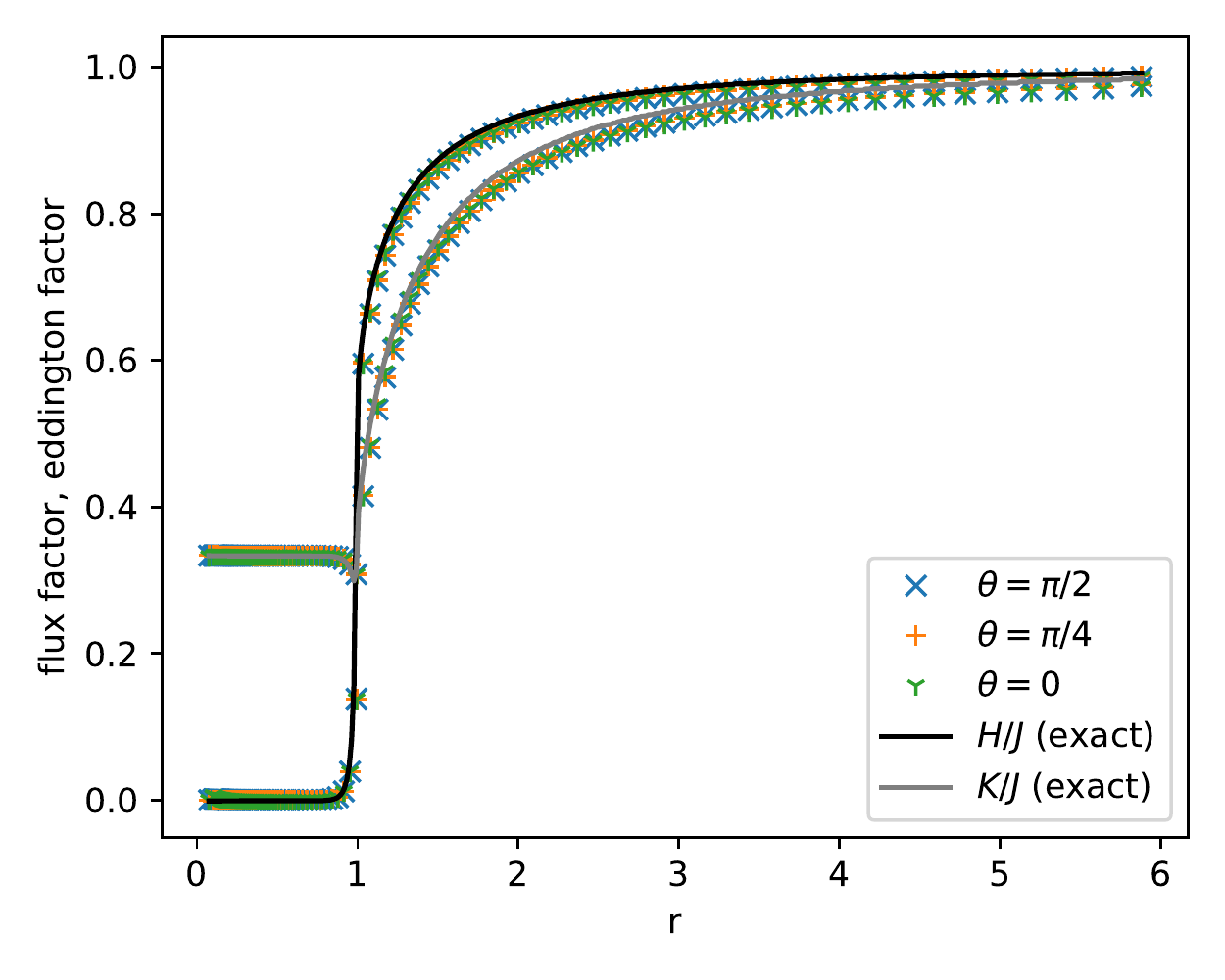}
    \hfill
    \includegraphics[width=0.49\linewidth]{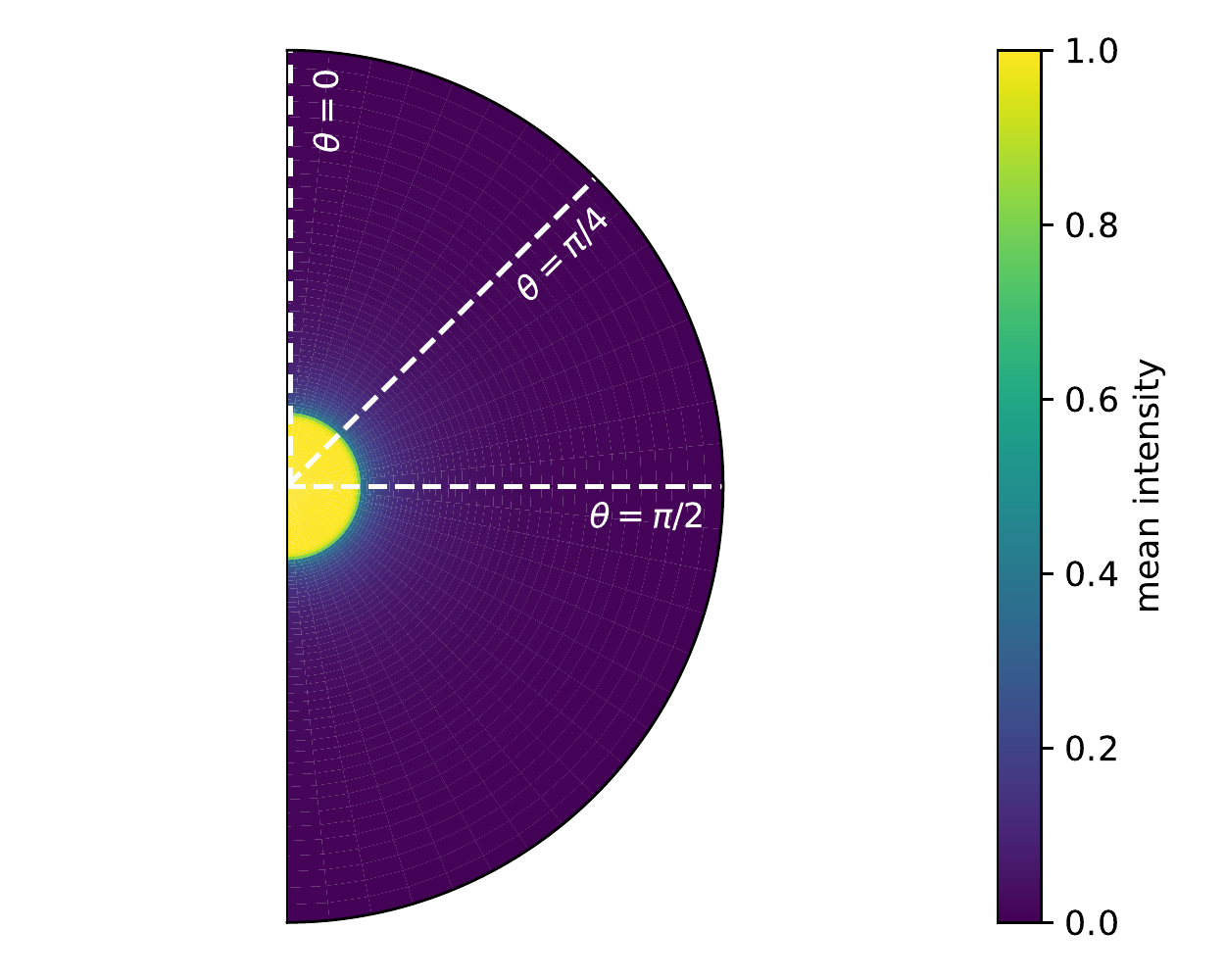}
    \caption{Left: Flux factor and Eddington factor along 3 rays (crosses) compared to the exact solutions (lines). The numerical solution along each ray is nearly identical, as expected. Right: Dashed lines the rays where the numerical solution is compared to the exact solution.}
    \label{fig:test_sphere_2d}
\end{figure*}

We repeat the classical optically thick radiating sphere problem on the 2D grid (Figure \ref{fig:test_sphere_2d}). While the initial conditions are spherically symmetric, it is worthwhile performing this test in axisymmetry given that cell volumes and areas differ along the $\theta$-direction on a spherical polar grid.  We use $N_r=100$, $N_\theta=32$, $N_\mu=17$, and $N_\Phi=32$.

We again find excellent agreement with the analytic solution, with a small error at the poles. Given the coordinate singularity present at the poles, this is unsurprising, and the scheme performs remarkably well despite this fact.

\subsubsection{Radiating disk}
Many two-moment methods have difficulty handling intersecting beams of radiation inasmuch as they produce a solution in which the beams erroneously interact. This limitation arises because the individual directions of propagation cannot be resolved with the
knowledge of
the moments $J$ and $H$ alone. To demonstrate the multi-dimensional capabilities of our code, we perform a test using very asymmetric initial conditions: a spatially thin disk radiating isotropically into a vaccuum. To reproduce the correct intensity at any point above the disk, the scheme must be able to accurately handle intersecting beams. Near the surface of the disk, the flux vector is misaligned with the radial coordinate, the direction with finest angular resolution in momentum space. This poses an additional challenge to accurately resolving the transition to forward-peaked radiation.

This problem is run in the domain $0<\theta<\frac{\pi}{2}$, with a vacuum (i.e. $\ka=0$) everywhere, and an isotropic distribution imposed along the boundary $\theta=\frac{\pi}{2}$, $-1<r<1$ simulating an optically thick,
but geometrically thin disk with $f_\mathrm{eq}=1$. We use $N_r=100$, $N_\theta=32$, $N_\mu=17$, and $N_\Phi=32$. 
We found that such a high resolution in angle is required
to obtain an acceptable numerical solution for this particular
test problem, but note that one cannot draw conclusions about
resolution requirements for a less extreme problem geometry
as typically encountered in core-collapse supernovae (with
the exception of collapsar disks).
Since the disk is optically thick, the exact solution at any point can be found by integrating the equilibrium distribution function over the projected surface of the disk
\begin{equation}
    \mathcal{J}(x,z) = \frac{1}{4\pi} \int_0^1 \int_0^{2\pi} \frac{rz\,\ud\phi\,\ud r}{\left( \left(x-r\cos\phi\right)^2 + z^2 + \left(r\sin\phi\right)^2 \right)^{3/2}}
\end{equation}
where $x$ and $z$ are the radius and height in \emph{cylindrical} coordinates.

\begin{figure*}
    \includegraphics[width=0.49\linewidth]{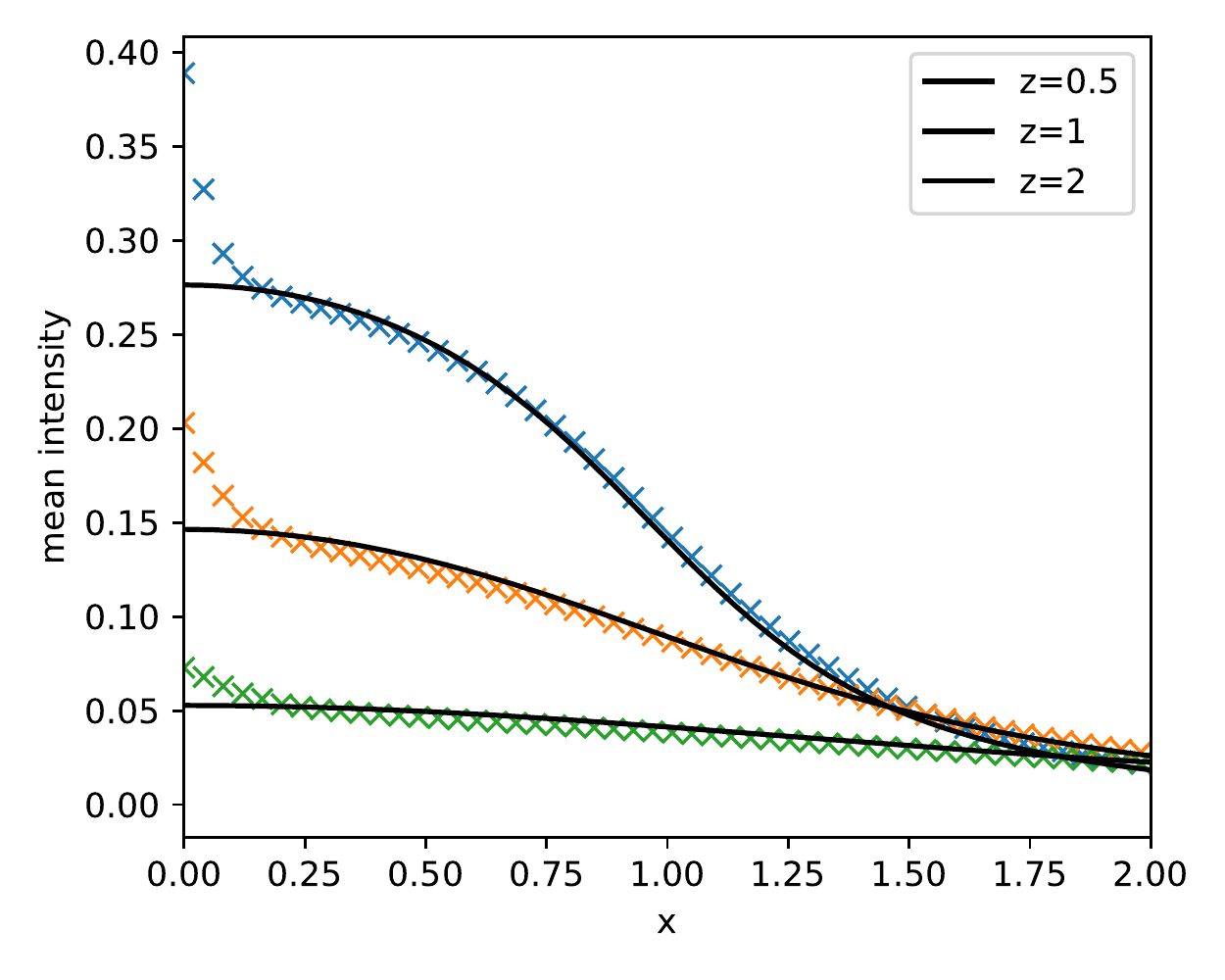}
    \hfill
    \includegraphics[width=0.49\linewidth]{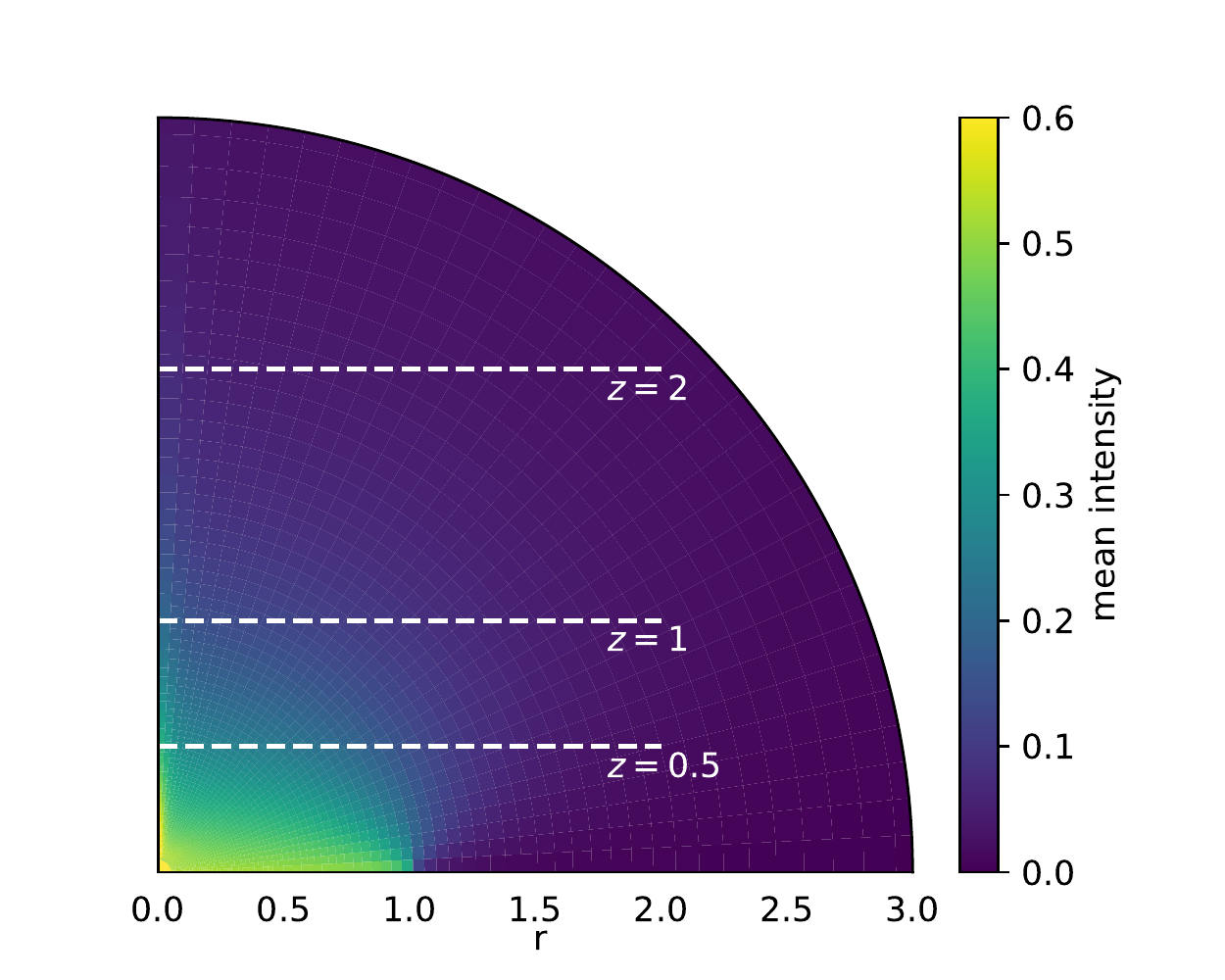}
    \caption{Left: Mean intensity as a function of distance from the center at three heights above the disk calculated using our scheme (crosses), compared to the exact solution (line). Right: Dashed lines indicate heights at which the intensity is compared to the exact solution.}
    \label{fig:test_disk}
\end{figure*}

We compare the mean intensity calculated using our scheme against the exact solution at constant heights above the disk (Figure \ref{fig:test_disk}), and find reasonable agreement, aside from an artefact at $x=0$ arising from the singularity of the spherical polar grid. The error is confined to the innermost three angular zones closest to the pole and could
conceivably be eliminated using a range of numerical treatments. In future 3D simulations, one may resort to singularity-free grids, such as the Yin-Yang overset grid \citep{2004GGG.....5.9005K}, so the axis artefact is not a major obstacle against generalising our
new algorithm to three dimensions. 

\section{Summary and Conclusions} \label{sec:summary}

Despite recent progress in developing robust and efficient
approximations for neutrino transport in multi-dimensional
core-collapse supernova models, simulations will ultimately
have to deal with the full kinetic equations and drop
the approximations inherent in diffusion and two-moment closure schemes. In this paper, we have introduced a new, locally implicit scheme for solving the full general relativistic Boltzmann transport equation in the context of neutrino transport in core-collapse supernovae.

Our scheme uses the discrete-ordinate method
to discretise the six-dimensional phase space into volume elements. We eliminate the need to  
implement angle and energy derivatives associated with geometric rotations and relativistic effects 
by using a semi-Lagrangian interpolation method to account for the frame transformations between
adjacent cells. The propagation of neutrinos in a relativistic spacetime is 
taken care of by applying appropriate Hamiltonian ``kicks'' to advected neutrinos.
Although the algorithm can accommodate arbitrary spacetime metrics, we limit
ourselves to a conformally flat metric
\citep{2008IJMPD..17..265I,1996PhRvD..54.1317W,2009PhRvD..79b4017C}
in our current work.
The scheme guarantees neutrino number conservation by construction,
and also has very good energy conservation properties. 
Neutrino fluxes between cells are computed using a semi-implicit Lax-Wendroff discretisation of the Boltzmann equation, which is consistent with the diffusion limit. In our scheme, we discretise the collision integral implicitly to maintain stability. By discretising the advection terms explicitly, we avoid a global coupling in space, which obviates the need for a complicated infrastructure
for global iterative solves of huge non-linear systems, and makes parallel scaling of massive simulations easier.
The simplicity of the algorithm and the purely local treatment of the collision integral
will also facilitate future extensions to the code (e.g., additional
interaction processes, generalisation to the quantum kinetic equations).

We have shown, using a suite of analytically solvable tests in spherical symmetry, that our scheme is accurate both in the diffusive regime and  the free-streaming regimes. We have also shown using tests in axisymmetry that our scheme is equally applicable in 2D, and argue that it will extend easily to 3D. We have demonstrated that our scheme, when coupled to the general relativistic hydrodynamics solver \textsc{CoCoNuT}, can be used to model the collapse and accretion
phase of core-collapse supernovae in spherical symmetry. Currently, the coupled neutrino
hydrodynamics code uses a reduced set of neutrino interactions.
We have not yet included non-isoenergetic scattering (NIS) in our collision integral, which 
critically affects the dynamics of the collapse phase \citep{1985ApJS...58..771B,1993ApJ...410..740M}, treat Bremsstrahlung as a one-particle reaction, and omit the pair process and neutrino pair conversion \citep{2003ApJ...587..320B}. The collapse test also revealed some residual
problems with entropy and lepton fraction advection errors in the innermost zones during
the end of the collapse phase, which can likely be fixed using a more careful treatment
of the origin of the spherical polar grid. A more detailed comparison with other
 neutrino transport codes in the vein of
\citet{2005ApJ...620..840L,2012ApJ...756...84M,2017ApJ...847..133R,2018JPhG...45j4001O,2018MNRAS.481.4786J} will be carried out once the required microphysics is in place.

Computer time requirements for our algorithm will increase significantly once the
collision integral is treated in its most general form.
In order to maximise the accuracy and efficiency of our new Boltzmann scheme and facilitate
its application in multi-dimensional supernova simulations, one can consider a number
of improvements to the algorithm.
A disadvantage of discrete ordinate schemes is that they require high angular resolution to resolve the distribution as $\mu\rightarrow 1$ in the free-streaming limit. Although our scheme already performs quite well because we have used $\mu=1$ as a collocation point, our mapping method lends itself to easy implementation of an adaptive grid in momentum space proposed by \cite{1999A&A...344..533Y}. In practice, however, this may yield only a minimal improvement in core-collapse models, since the forward-peaked regions are far outside the region of dynamical influence, but an adaptive grid in angle and/or energy might allow us to achieve
comparable results with significantly lower resolution in momentum space.

More substantial savings may be gained by introducing local timestepping.
In our implementation, we have used global timesteps, which are limited by the small zones at the center of the grid. Since the timestep can be much larger in the larger outer zones, a local timestepping method may provide computational cost savings. Furthermore, the timestep can be increased above the CFL limit in diffusive zones without compromising stability. 
We will also explore numerical techniques to optimise the implicit solution of
the neutrino source terms for complicated interaction kernels.

A major uncertainty in the treatment of neutrinos in core-collapse supernovae
comes from the effects of neutrino flavour conversion. 
While the classical Mikheyev-Smirmov-Wolfenstein (MSW) effect 
\citep{1978PhRvD..17.2369W,1985YaFiz..42.1441M} is significant at densities of 
$10^0\texttt{-}10^3\,\mathrm{g}/\mathrm{cm}^3$, and only influences the observed neutrino 
signal rather than the dynamics in the supernova core,
flavour conversion due to $\nu$-$\nu$ interactions 
\citep[e.g.][]{1992PhLB..287..128P,2010ARNPS..60..569D,2016NCimR..39....1M}
may occur deeper in the core and have tangible repercussions on the dynamics
of the pre-explosion and explosion phase, and on supernova nucleosynthesis.
 In particular, multi-angle effects \citep[e.g.][]{2005PhRvD..72d5003S,2009PhRvD..79j5003S,2012PhRvL.108z1104C,2017JCAP...02..019D}
could result in fast flavour conversion already in the region of the neutrinosphere.
Flavour conversion cannot be consistently implemented in supernova simulations yet.\footnote{There are first ideas to use solvers for the quantum kinetic
equations for more than post-processing \citep{2019arXiv191004172S}, but these
are not fully convincing and far from achieving a consistent coupling of the hydrodynamics
with the quantum kinetic equations}.

Since the detailed angular distribution of neutrinos can be critical in the flavour
conversion problem, and since the validity of closure approximations in the
quantum kinetic regime is doubtful, a multi-dimensional Boltzmann treatment is
an essential stepping stone towards a consistent inclusion of flavour conversion
in supernova simulations in the long term.
Incorporating flavour conversion will entail replacing the (flavour-dependent)
distribution function with a density matrix that obeys the equations of quantum
kinetic theory \citep[e.g.][]{1993NuPhB.406..423S,2008PhRvD..78h5017C,2015IJMPE..2441009V,2019PhRvD..99l3014R}. Although considerable work is still needed to understand the complicated instabilities
in the non-linear flavor conversion problem, 
our scheme provides a suitable framework for implementing the relevant macroscopic equations.
While a consistent treatment of flavour conversion in supernovae is not yet on the horizon
and will require prodigious interdisciplinary efforts, we believe there may be
considerable use for a conceptually simple, locally implicit algorithm for neutrino kinetics
as outlined in this paper.

\section*{Acknowledgements}
We thank Alexander Heger for useful discussions. CC was supported by an Australian Government Research Training Program (RTP) Scholarship. This work was supported by the Australian Research Council through ARC Future Fellowship FT160100035. We acknowledge CPU time on OzSTAR funded by Swinburne University and the Australian Government. This research was undertaken with the assistance of resources obtained via NCMAS and ASTAC  from the National Computational Infrastructure (NCI), which is supported by the Australian Government and was supported by resources provided by the Pawsey Supercomputing Centre with funding from the Australian Government and the Government of Western Australia. 

\appendix

\section{Interface fluxes in the diffusion limit} \label{app:diffusion}
The distribution function can be expressed as an expansion of its moments
\begin{equation}
    f = \mathcal{J} + \mu\mathcal{H} + \cdots
\end{equation}
to solve the  Boltzmann equation, which can be expressed in planar geometry as
\begin{equation}
    \frac{\pd f}{\pd t} + \mu \frac{\pd f}{\pd z} = \ka \left(\mathcal{J}_\mathrm{eq} - f\right) + \ks \left(\mathcal{J} - f\right).
\end{equation}
The equation for the first moment becomes
\begin{equation}
    \frac{\pd \mathcal{H}}{\pd t} + \frac{\pd \mathcal{K}}{\pd z} = - \left(\ka + \ks\right)\mathcal{H}.
\end{equation}
In the diffusion limit, $\frac{\pd\mathcal{H}}{\pd t}\rightarrow 0$, thus we obtain
\begin{equation}
    \mathcal{H} = -\frac{1}{\ka+\ks} \frac{\pd \mathcal{K}}{\pd z}.
    \label{eq:diffusion}
\end{equation}
We seek a discretisation of the Boltzmann equation that approaches the same limit of $\mathcal{H}$ for the numerical fluxes. In the Lax-Wendroff scheme, the fluxes at the interface between $i$ and $i+1$, labelled $i+\hf$, for a particular direction $\mu_j$, are
\begin{equation}
    f_{i+\hf}^{n+\hf} = f_{i+\hf}^n + \frac{\dt}{2} \left(\mu_j\frac{f_i^n - f_{i+1}^n}{\Delta z}\right) + \frac{\dt}{2}\ka\left(\mathcal{J}_\mathrm{eq} - f_{i+\hf}^{n+\hf}\right) + \frac{\dt}{2}\ks\left(\mathcal{J} - f_{i+\hf}^{n+\hf}\right).
\end{equation}
For optically thick cells, $\frac{\Delta t}{2} \left(\ka + \ks\right) \gg 1$, thus
\begin{equation}
    \mu\frac{f_i^n - f_{i+1}^n}{\Delta z} = \ka\left(\mathcal{J}_\mathrm{eq} - f_{i+\hf}^{n+\hf}\right) + \ks\left(\mathcal{J} - f_{i+\hf}^{n+\hf}\right).
\end{equation}
Multiplying by $\mu$, integrating over all directions $j$, and simplifying gives
\begin{align}
    \frac{\sum_j \mu_j^2f_{i+1}^n\Delta\mu_j - \sum_j \mu_j^2f_i^n\Delta\mu_j}{\Delta z} &= -\left(\ka + \ks\right) \times \sum_j \mu_j f_{i+\hf}^{n+\hf} \Delta\mu_j \\
    \frac{\Delta\mathcal{K}_{i+\hf}^n}{\Delta z} &= -\left(\ka + \ks\right) \mathcal{H}_{i+\hf}^{n+\hf},
\end{align}
which is the discretised form of Equation~(\ref{eq:diffusion}) at cell interfaces.

\bibliographystyle{mnras}
\bibliography{references}

\end{document}